\documentclass[12pt]{article}
\usepackage[latin1]{inputenc}
\usepackage[dvips]{graphicx,color}
\usepackage{amssymb}
\usepackage{hyperref}
\usepackage{bibspacing}

\def\beq{\begin{equation}}
\def\eeq{\end{equation}}
\def\eq{\end{equation}}
\def\ba{\begin{eqnarray}}
\def\ea{\end{eqnarray}}

\def\centeron#1#2{{\setbox0=\hbox{#1}\setbox1=\hbox{#2}\ifdim
\wd1>\wd0\kern.5\wd1\kern-.5\wd0\fi
\copy0\kern-.5\wd0\kern-.5\wd1\copy1\ifdim\wd0>\wd1
\kern.5\wd0\kern-.5\wd1\fi}}
\def\ltap{\;\centeron{\raise.35ex\hbox{$<$}}{\lower.65ex\hbox{$\sim$}}\;}
\def\gtap{\;\centeron{\raise.35ex\hbox{$>$}}{\lower.65ex\hbox{$\sim$}}\;}
\def\gsim{\mathrel{\gtap}}
\def\lsim{\mathrel{\ltap}}

\newcommand{\captionfonts}{\small}
\makeatletter \long\def\@makecaption#1#2{
  \vskip\abovecaptionskip
  \sbox\@tempboxa{{\captionfonts #1: #2}}
  \ifdim \wd\@tempboxa >\hsize
    {\captionfonts #1: #2\par}
  \else
    \hbox to\hsize{\hfil\box\@tempboxa\hfil}
    \fi
  \vskip\belowcaptionskip}
\makeatother

\newcommand{\newc}{\newcommand}
\newc{\qbar}{{\overline q}}
\newc{\Kahler}{K\"ahler }
\newc{\deltaGS}{\delta_{\rm GS}}
\newc{\bsg}{B\rightarrow X_s\gamma}
\newc{\Bmumu}{B_s\rightarrow \mu^+ \mu^-}
\newc{\MSusy}{m_{\tilde{t}}}
\newc{\HEWSB}{H^{\textrm{\tiny{EWSB}}}}
\newc{\cossqbma}{\cos^2(\beta-\alpha)}
\newc{\sinsqbma}{\sin^2(\beta-\alpha)}
\newc{\HSM}{H_{\textrm{\tiny{SM}}}}
\newc{\MSusyMin}{m_{\tilde{t},\textrm{\tiny{min}}}}
\newc{\MGUT}{M_{\mbox{\scriptsize{GUT}}}}

\begin{document}
\def\pplogo{\vbox{\kern-\headheight\kern -29pt
\halign{##&##\hfil\cr&{\ppnumber}\cr\rule{0pt}{2.5ex}&\ppdate\cr}}}
\makeatletter
\def\ps@firstpage{\ps@empty \def\@oddhead{\hss\pplogo}
  \let\@evenhead\@oddhead
}
\def\maketitle{\par
 \begingroup
 \def\thefootnote{\fnsymbol{footnote}}
 \def\@makefnmark{\hbox{$^{\@thefnmark}$\hss}}
 \if@twocolumn
 \twocolumn[\@maketitle]
 \else \newpage
 \global\@topnum\z@ \@maketitle \fi\thispagestyle{firstpage}\@thanks
 \endgroup
 \setcounter{footnote}{0}
 \let\maketitle\relax
 \let\@maketitle\relax
 \gdef\@thanks{}\gdef\@author{}\gdef\@title{}\let\thanks\relax}
\makeatother

%%%%%%%%%%%%%%%%%%%%%%%%%%%%%%%%%%%%%%%%%%%%%%%%%%%%%%%%%%%%%%

\setcounter{page}0
\def\ppnumber{\vbox{\baselineskip14pt
%\hbox{hep-ph/0000000}
}}
\def\ppdate{RUNHETC-2007-14} \date{}

\author{\\Rouven Essig\footnote{rouven@physics.rutgers.edu},
Jean-Fran\c{c}ois Fortin\footnote{jffor27@physics.rutgers.edu} \\
[7mm]
{\normalsize NHETC, Department of Physics and Astronomy,}\\
{\normalsize Rutgers University, Piscataway, NJ 08854, U.S.A.}\\}

\title{\bf \LARGE The Minimally Tuned Minimal Supersymmetric Standard Model} \maketitle \vskip 2cm

\begin{abstract} \normalsize
\noindent  The regions in the Minimal Supersymmetric Standard Model with
the minimal amount of fine-tuning of electroweak symmetry breaking
are presented for general messenger scale.
No a priori relations among the soft supersymmetry breaking
parameters are assumed and fine-tuning is minimized with
respect to all the important parameters which affect electroweak
symmetry breaking.
The superpartner spectra in the minimally tuned region of parameter
space are quite distinctive with large stop mixing at the low scale and
negative squark soft masses at the high scale.
The minimal amount of tuning increases enormously for a Higgs mass
beyond roughly 120 GeV.
\end{abstract}
\bigskip

\newpage

\tableofcontents

%%%%%%%%%%%%%%%%%%%%%%%%%%%%%%%%%%%%%%%%%%%%%%%%%%%%%%%%%%%%%%%%%%%%%%%%
%%%%%%%%%%%%%%%%%%%%%%%%%%%%%%%%%%%%%%%%%%%%%%%%%%%%%%%%%%%%%%%%%%%%%%%%
%%%%%%%%%%%%%%%%%%%%%%%%%%%%%%%%%%%%%%%%%%%%%%%%%%%%%%%%%%%%%%%%%%%%%%%%
%%%%%%%%%%%%%%%%%%%%%%%%%%%%%%%%%%%%%%%%%%%%%%%%%%%%%%%%%%%%%%%%%%%%%%%%
%%%%%%%%%%%%%%%%%%%%%%%%%%%%%%%%%%%%%%%%%%%%%%%%%%%%%%%%%%%%%%%%%%%%%%%%
%%%%%%%%%%%%%%%%%%%%%%%%%%%%%%%%%%%%%%%%%%%%%%%%%%%%%%%%%%%%%%%%%%%%%%%%

\section{Introduction}\label{Sec:Intro}

The Minimal Supersymmetric Standard Model (MSSM) is a well-motivated
candidate for physics beyond the Standard Model (SM).
The gauge couplings within the MSSM unify to within a few percent at the
grand unified theory (GUT) scale, $\MGUT\simeq 2\times 10^{16}$ GeV, and
the lightest supersymmetric particle is a good dark matter candidate
provided that R-parity is conserved.
Supersymmetry (SUSY) can also naturally stabilize the hierarchy between
the electroweak (EW) and the GUT or Planck scale.
It does this by providing a radiative mechanism for electroweak symmetry
breaking (EWSB) where large quantum fluctuations of the scalar top squarks
due to the large Yukawa coupling destabilize the origin of the Higgs potential.
In much of the MSSM parameter space this quite naturally leads to the
right EWSB scale, as long as the soft SUSY breaking parameters lie near it.

The absence of any direct experimental evidence from collider searches for
the MSSM scalar particles and the Higgs boson has, however, ruled out
significant regions in the MSSM parameter space.
Indirect evidence from EW precision measurements
and searches for flavor changing neutral currents, CP violating effects and rare
decays has not been forthcoming either, providing additional
severe constraints.
As a result the soft SUSY breaking parameters must lie well
above the EW scale in order to satisfy the experimental constraints,
especially the constraints on the Higgs mass from the results of the CERN LEP collider
($m_h\gtrsim 114.4$ GeV \cite{ALEPH:2006cr}).

Soft SUSY breaking parameters well above the EW scale reintroduce a
small hierarchy and require some fine-tuning (FT)
among the SUSY parameters in order to obtain EWSB
\cite{Ellis:1986yg}-\nocite{Barbieri:1987fn,deCarlos:1993yy,deCarlos:1993ca,Anderson:1994dz,Ciafaloni:1996zh,Chankowski:1997zh,Agashe:1997kn,Chan:1997bi,Wright:1998mk,Kane:1998im,BasteroGil:1999gu,Casas:2003jx,Casas:2004uu,Casas:2004gh,Schuster:2005py,Dermisek:2005ar,Dermisek:2006ey,Casas:2006bd,Kobayashi:2006fh}\cite{Athron:2007ry}.
This is usually referred to as the supersymmetric little hierarchy
problem.

Different choices for the soft SUSY breaking parameters lead to
different amounts of FT.
This paper presents the minimally tuned MSSM (or MTMSSM), i.e.
the MSSM parameter region that has the least model-independent
FT of EWSB.
Model-independent means that no relations are assumed between the soft SUSY
breaking parameters at the scale at which they are generated
(which will be referred to as the messenger scale).
Rather, each of them is taken to be an independent parameter
which is free at the messenger scale, and which therefore can
contribute to the total FT of the EWSB scale.
The messenger scale itself is varied between 2 TeV and $\MGUT$
and the effect of this on the minimal FT is discussed
(see also \cite{Casas:2003jx,Casas:2004uu}).

In Section \ref{Sec:EWSB}, EWSB in the MSSM will be reviewed.
Section \ref{Sec:Tuning Measure} discusses the tuning measure used in this
paper.
The parameters taken to contribute to the tuning are $|\mu|^2$, $m_{H_u}^2$,
the gaugino masses $M_1$, $M_2$ and $M_3$, the stop soft masses $m^2_{\tilde{t}_L}$
and $m^2_{\tilde{t}_R}$, and the stop soft trilinear coupling $A_t$.

Section \ref{Sec:MMIT} contains some of the main results.
The low- and high-scale MSSM spectrum which leads to the least
model-independent FT is found.
This is done for various messenger scales by numerically minimizing the
FT expression subject to constraints on the Higgs, stop, and gaugino masses.
The results are then motivated analytically.
The least FT is found to be about 5$\%$ if the messenger scale coincides with the GUT scale.
An important feature of the least FT region is negative stop soft masses
at the messenger scale (first pointed out in \cite{Dermisek:2006ey}).
Even for messenger scales as low as 2 TeV, the stop soft masses are tachyonic at
the messenger scale (threshold effects in the RG-running were neglected throughout).
This does not lead to any problems with charge and/or color breaking minima.
Another feature of the least FT region is that the trilinear stop soft
coupling, $A_t$, is negative and lies near ``natural'' maximal mixing, i.e.
$A_t\simeq-2\MSusy$, where $\MSusy$ is the average of the two stop soft masses.
This value for $A_t$ maximizes the radiative corrections to $m_h$.
The large stop mixing leads to a sizeable splitting between the two stop
mass eigenstates.
Moreover, the gluino mass, $M_3$, is much smaller than the wino mass, $M_2$,
at the high scale.
The wino mass, in turn, is much smaller than the bino mass $M_1$.
Phenomenological consequences of the low-scale spectrum are briefly summarized.

Section \ref{Sec:FT_vs_mh} contains the rest of the main results of the paper.
The FT is minimized as a function of the lower bound on the Higgs mass (with
the messenger scale set to $\MGUT$).
Although the numerical minimization procedure contains the dominant one-loop
expression for $m_h$ as a constraint, the resulting least FT spectra
are used to calculate $m_h$ more accurately with the program $\mathtt{FeynHiggs}$
\cite{Frank:2006yh,Degrassi:2002fi,Heinemeyer:1998np,Heinemeyer:1998yj,Heinemeyer:2007aq}.
The result is a plot of the minimal FT as a function of $m_h$.
There are several striking features of this plot.
First of all, for $m_h$ larger than a certain value, the FT increases very rapidly.
Secondly, around this $m_h$, the value of $A_t$ in the least FT region makes a sudden
transition from lying near $-2\MSusy$ to lying near $+2\MSusy$.
The third striking feature is that this value of $m_h$ is surprisingly low.
The precise value is only slightly dependent on the parameters in the Higgs sector
and can be taken to lie around 120 GeV.
The upshot of this analysis is that although the MSSM right now is already fine-tuned
at least at about the $5\%$ level (if the messenger scale equals the GUT scale), there is not much
room left for the Higgs mass to increase before the FT becomes exponentially worse.

Section \ref{Sec:Conclusion} contains a summary of the results and the conclusions.
Appendix \ref{Sec:Master Formula} reviews the semi-numerical solutions
of the MSSM one-loop renormalization - group (RG) equations.
These are used to calculate the expression for the FT employed in this paper.
Appendix \ref{Sec:FTComps} contains a list of expressions for the FT with
respect to various parameters.

\section{Electroweak Symmetry Breaking}\label{Sec:EWSB}

In the Higgs decoupling limit of the MSSM, the lower bound on the mass
of the lighter CP-even Higgs mass eigenstate $h$ coincides with the 114.4 GeV bound
on the mass of the SM Higgs boson \cite{ALEPH:2006cr}.
The mass of $h$ may be approximated by
\beq\label{Eqn:Higgs mass}
m_h^2 \simeq m_Z^2\,\cos^2 2\beta + \frac{3}{4\pi^2}\,\frac{m_t^4}{v^2}\,\left[\log\frac{\MSusy^2}{m_t^2} + \frac{X_t^2}{\MSusy^2}\,\left(1-\frac{X_t^2}{12\MSusy^2}\right)\right]
\eeq
which, in addition to the tree-level Higgs mass, includes the dominant one-loop quantum
corrections coming from top and stop loops \cite{Okada:1990vk,Ellis:1990nz,Haber:1990aw,Haber:1996fp,Carena:1995bx,Carena:1995wu}.
Here $m_t$ is the top mass, $\MSusy^2$ is the arithmetic mean of
the two squared stop masses and $v=2m_W/g\simeq 174.1$ GeV
where $g$ is the $SU(2)$ gauge coupling and $m_W$ is the mass of the $W$-boson.
Furthermore, equation (\ref{Eqn:Higgs mass}) assumes $\MSusy$ $\gg$ $m_t$.
The stop mixing parameter is given by $X_t=A_t - \mu\cot\beta$
($\simeq$ $A_t$ for large $\tan\beta$), where $A_t$ denotes the
stop soft trilinear coupling and $\mu$ is the supersymmetric
Higgsino mass parameter.
The first term in equation (\ref{Eqn:Higgs mass}) is the tree-level
contribution to the Higgs mass.  The first term in square brackets comes
from renormalization group running of the Higgs quartic coupling
below the stop mass scale and vanishes in the limit of exact
supersymmetry.
It grows logarithmically with the stop mass.
The second term in square brackets is only present for non-zero stop mixing and comes from
a finite threshold correction to the Higgs quartic coupling at the
stop mass scale.
It is independent of the stop mass for fixed
$X_t/\MSusy$, and grows as $(X_t/\MSusy)^2$ for small
$X_t/\MSusy$.

Equation (\ref{Eqn:Higgs mass}) implies a combination of three things
which are required to satisfy the bound on $m_h$, namely a large tree-level
contribution, large stop masses and large stop mixing.
A large tree-level contribution to $m_h$ requires $\tan\beta$ to be at least
of a moderate size ($\gtrsim 5-10$).
Although the stop masses must be rather large, their lower bound is very sensitive
to the size of the stop mixing, with larger mixing allowing for much smaller stop masses
(see \cite{Essig:2007vq} for a recent study on this).
The reason for this sensitive dependence is due to the Higgs mass
depending logarithmically on the stop masses in contrast to the polynomial
dependence on the stop mixing.

The soft masses are not only directly constrained from the LEP Higgs bounds but also indirectly
by constraints on flavor changing neutral currents, electroweak precision measurements and CP-violation.
Besides these, however, the Higgs sector parameters are also constrained by requiring that the
electroweak symmetry is broken.
This leads to the following two tree-level relations at the low scale
\beq\label{Eqn:EWSBsin2beta}
\sin 2\beta = \frac{2 m_{12}^2}{m_{H_u}^2 + m_{H_d}^2 + 2|\mu|^2} = \frac{2m_{12}^2}{m_A^2}
\eeq
\beq\label{Eqn:EWSBmZ1}
\frac{m_Z^2}{2} = -|\mu|^2 + \frac{m_{H_d}^2 - m_{H_u}^2 \tan^2\beta}{\tan^2\beta-1},
\eeq
where $m_A$ is the CP-odd Higgs mass, and $\beta$ is determined from the ratio
of the two vacuum expectation values $v_u\equiv\langle{\rm Re}
(H_u^0)\rangle$ and $v_d\equiv\langle{\rm Re} (H_d^0)\rangle$ as
$\tan\beta = v_u/v_d$.
The masses $m_{H_u}^2$, $m_{H_d}^2$ and $m_{12}^2$ are the three soft mass parameters
in the MSSM Higgs sector.
For a given value of $\tan\beta$, $m_{12}^2$ may be eliminated in favor of $m_A^2$ with
equation (\ref{Eqn:EWSBsin2beta}).
Equation (\ref{Eqn:EWSBmZ1}) gives an expression for $m_Z^2$ in terms of the supersymmetric
mass parameter $\mu$ and the soft masses $m_{H_u}^2$ and $m_{H_d}^2$.
Since $\tan\beta$ should be sizeable, the contribution from $m_{H_d}^2$
to the expression for $m_Z^2$ may be neglected and (\ref{Eqn:EWSBmZ1}) simplifies to
\beq\label{Eqn:EWSBmZ2}
m_Z^2 = -2|\mu|^2 - 2m_{H_u}^2% + \mathcal{O}\left(\frac{2(m_{H_d}^2-m_{H_u}^2)}{\tan^2\beta}\right).
.
\eeq
Close to the Higgs decoupling limit, $m_A$ is relatively large.
However, since $|\mu|^2,m_{H_u}^2\sim\mathcal{O}(m_Z^2)$ to avoid large cancellations, $m_A$ may not be too
large, otherwise $m_{H_d}^2$ would also be sizeable
and equation (\ref{Eqn:EWSBmZ2}) would break down (unless the value of
$\tan\beta$ is increased accordingly).
By choosing $\tan\beta=10$ and $m_A=250$ GeV in the numerical analysis throughout, equation
(\ref{Eqn:EWSBmZ2}) holds to a very good approximation.

Equation (\ref{Eqn:EWSBmZ2}) holds at tree-level, and although quantum
corrections may add $\mathcal{O}$(10 GeV) to the
right hand side of (\ref{Eqn:EWSBmZ2}), this has negligible
impact on the amount of fine-tuning to be discussed below.

\begin{figure}[t]\begin{center}\includegraphics[scale=0.5]{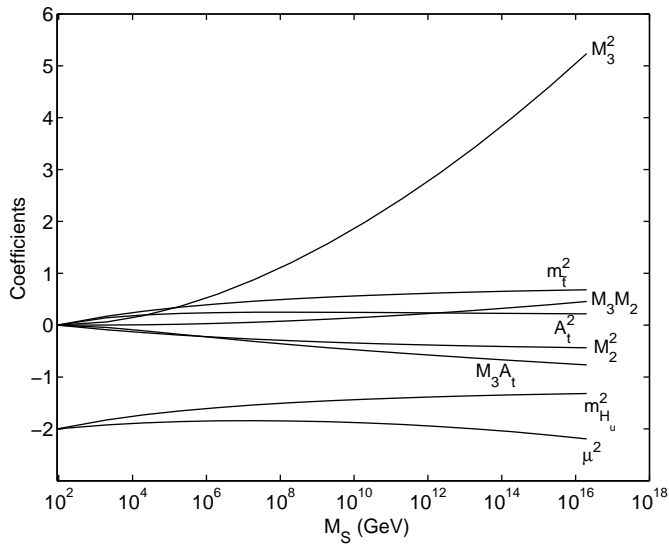}
\caption{The coefficients $c_{ij}$ defined in equation (\ref{Eqn:mZsq generic})
for $\tan\beta=10$ as a function of the messenger scale $M_S$.
\label{Fig:coefficients}}
\end{center}\end{figure}

The parameters $m_{H_u}^2$ and $|\mu|^2$ in equation (\ref{Eqn:EWSBmZ2}) are evaluated
at the scale $m_Z$.
Since the fine-tuning of EWSB is a measure of the sensitivity of
some low-scale EWSB parameter (usually taken to be $m_Z^2$) to a change in
high-scale input parameters, $|\mu|^2$ and %the soft supersymmetry breaking mass
$m_{H_u}^2$ need to be evolved to a high scale using their RG equations.
Under RG running many of the soft parameters mix, and as a result of this mixing,
the expression for $m_Z^2$ in terms of parameters that are evaluated
at the messenger scale $M_S$ differs significantly from the simple form
given in (\ref{Eqn:EWSBmZ2}).
The RG-equations may be integrated (see Appendix \ref{Sec:Master Formula})
and the expression for $m_Z^2$ may generically be written as \cite{Ibanez:1983di,Carena:1996km}
\beq\label{Eqn:mZsq generic}
m_Z^2 = \sum_{i,j} c_{ij}(\tan\beta,M_S)\,m_i(M_S)\,m_j(M_S).
\eeq
For moderate and not too large values of $\tan\beta$ with an appropriate $m_A$, the simplified
expression for $m_Z^2$ is applicable (equation (\ref{Eqn:EWSBmZ2})) and contributions from the
bottom/sbottom and tau/stau sectors may still be neglected.
The most important parameters appearing in (\ref{Eqn:mZsq generic}) then are $\mu^2$, $m_{H_u}^2$,
the gaugino masses $M_1$, $M_2$ and $M_3$, the stop soft masses $m^2_{\tilde{t}_L}$
and $m^2_{\tilde{t}_R}$, and the stop soft trilinear coupling $A_t$.
The coefficients $c_{ij}$ depend on $\tan\beta$ and the messenger scale $M_S$.
The most important coefficients are shown in Figure \ref{Fig:coefficients} for $\tan\beta = 10$
as a function of $M_S$.

At the scale $m_Z$, the coefficients of $m_{H_u}^2$ and $\mu^2$ are $-2$ while the coefficients
of the other soft parameters are zero in agreement with equation (\ref{Eqn:EWSBmZ2}).
Since $\mu^2$ is a supersymmetric parameter, it gets renormalized multiplicatively and
its RG evolution does not give rise to soft parameters (see equation (\ref{Eqn:mu})).
Figure \ref{Fig:coefficients} shows that the coefficient of $\mu^2$ does not vary much
and remains close to $-2$ all the way up to the GUT scale.
The RG evolution of $m_{H_u}^2$ to higher messenger scales, however, generates non-zero
coefficients for the other soft parameters.  The $\beta$-function of $m_{H_u}^2$,
\begin{equation}\label{Eqn:bmHu}
8\pi^2\beta_{m_{H_u}^2}=3\lambda_t^2(m_{H_u}^2+m_{\tilde{t}_L}^2+m_{\tilde{t}_R}^2
+|A_t|^2)-3g_2^2|M_2|^2-g_Y^2|M_1|^2-\frac{1}{2}g_Y^2S_Y,
\end{equation}
depends on the stop sector parameters $\{m_{\tilde{t}_L}^2,m_{\tilde{t}_R}^2,A_t\}$,
the wino and bino masses $M_2$ and $M_1$, and $S_Y\equiv\frac{1}{2}\mbox{Tr}(Y_im_i^2)$,
which thus get generated immediately under RG evolution.
The coefficients of $M_2$ and especially $M_1$ and $S_Y$ in (\ref{Eqn:bmHu}) are
small and lead to small coefficients in the expression for
$m_Z^2$ (\ref{Eqn:mZsq generic}).
Although $\beta_{m_{H_u}^2}$ does not explicitly depend on the gluino mass, a non-zero
coefficient for $M_3$ is generated indirectly since the stop sector $\beta$-functions
depend on $M_3$.
Moreover, $M_3$ appears with a large coefficient in these $\beta$-functions, and thus
the coefficient of $M_3$ in equation (\ref{Eqn:mZsq generic}) dominates after a few
decades of RG evolution.
For example, at a messenger scale of $M_S=M_{\mbox{\scriptsize{GUT}}}\equiv 2\times 10^{16}$ GeV,
the expression for $m_Z^2$ (for $\tan\beta$ = 10) is
\begin{eqnarray}\label{Eqn:MF1}
m_Z^2 & = & -2.19\, \hat{\mu}^2 -\, 1.32\,\hat{m}_{H_u}^2 +\, 0.68\, \hat{m}_{\tilde{t}_L}^2 +\, 0.68\, \hat{m}_{\tilde{t}_R}^2 +\, 5.24\, \hat{M}_3^2 -\, 0.44\, \hat{M}_2^2 \nonumber \\
      &   &  -\, 0.01 \hat{M}_1^2 +\, 0.22\, \hat{A}_t^2 - 0.77\, \hat{A}_t\,\hat{M}_3 -\, 0.17\, \hat{A}_t\, \hat{M}_2 -\, 0.02\, \hat{A}_t\, \hat{M}_1 \nonumber \\
      &   &  +\, 0.46\, \hat{M}_3\, \hat{M}_2 +\, 0.07\, \hat{M}_3\, \hat{M}_1 +\, 0.01\, \hat{M}_2\, \hat{M}_1 +\, 0.05\, \hat{S}_Y,
\end{eqnarray}
where the hatted parameters on the right-hand side are all evaluated at $M_S$.
This expression may be used to calculate the FT as discussed next.

\section{The Tuning Measure}\label{Sec:Tuning Measure}

A variety of tuning measures have been used in the literature
(a list of references has been provided in the Introduction).
Since the concept of fine-tuning (FT) is inherently subjective,
there is no absolute definition of a FT measure.
The most common definition of the sensitivity of an observable
$\mathcal{O}(\{a_i\})$ on a parameter $a_i$, denoted by
$\Delta(\mathcal{O},a_i)$, is given by \cite{Ellis:1986yg,Barbieri:1987fn}
\beq\label{Eqn:FTab}
\Delta(\mathcal{O},a_i) = \left|\frac{\partial\log\mathcal{O}}{\partial\log a_i}\right|
= \left|\frac{a_i}{\mathcal{O}}\,\frac{\partial \mathcal{O}}{\partial a_i}\right|\,.
\eeq
$\Delta(\mathcal{O},a_i)$ thus measures the percentage variation of the observable under a percentage
variation of the parameter.
A large value of $\Delta(\mathcal{O},a_i)$ signifies that a small change in the parameter leads to a large change in
the observable, and suggests that the observable is fine-tuned with respect to that parameter.
Assuming that the individual $\Delta(\mathcal{O},a_i)$ are uncorrelated,
they may be combined to form the FT measure
\beq\label{Eqn:FTdefngeneral}
\mathcal{F}(\mathcal{O}) =\sqrt{\sum_i \,\Big(\Delta(\mathcal{O},a_i)\Big)^2}.
\eeq

Of interest in this paper is to quantify the sensitivity of EWSB in the MSSM
on (soft) supersymmetric parameters at the messenger scale $M_S$.
To this end, the observable to consider is $m_Z^2$ as a function of the
supersymmetric Higgsino mass squared and the soft supersymmetry breaking
parameters, collectively denoted by $m_i^2(M_S)$ (in the FT measure, all parameters
are taken to have mass dimension two).
The sensitivity of $m_Z^2$ with respect to each parameter may be
calculated as in (\ref{Eqn:FTab}) with $\mathcal{O}=m_Z^2$, and the total FT of $m_Z^2$
on parameters evaluated at the messenger scale $M_S$ may be quantified by
\beq\label{Eqn:FTdefn}
\mathcal{F}(m_Z^2;M_S)=\sqrt{\sum_i \,\Big(\Delta\big(m_Z^2,m_i^2(M_S)\big)\Big)^2}.
\eeq
$\mathcal{F}(m_Z^2;M_S)$ may be interpreted as the length of
a ``fine-tuning vector'' with components $\Delta(m_Z^2,m_i^2(M_S))$.
This fine-tuning vector is formally a vector field defined by the
gradient of the scalar field $\log m_Z^2$, a function of $\log m_i^2$,
along surfaces of constant $\log m_Z^2$.

There are several possible drawbacks to this FT measure, see for
example \cite{Athron:2007ry,Athron:2007as}.
One of these is that the individual $\Delta(m_Z^2,m^2_i(M_S))$
are assumed to be uncorrelated.
Within a given model of supersymmetry breaking, there may be
relations among the parameters at the messenger scale.
This would imply that the FT vector is projected onto a
subspace, and the resulting FT is necessarily less.
In other words, the tuning of one parameter is correlated with
the tuning of another, so that the total FT should be less than
that given by (\ref{Eqn:FTdefn}).
Moreover, within a given model the values of the parameters at the
messenger scale may be restricted to certain ranges, whereas
(\ref{Eqn:FTdefn}) assumes that all values are equally likely.
However, no model for supersymmetry breaking will be assumed here.
Instead, the minimal FT will be found as a function of the
messenger scale $M_S$ assuming no relations or restrictions among
the high-scale input parameters.
For this ``model-independent'' tuning it is satisfactory to use the FT
measure (\ref{Eqn:FTdefn}).

Note that to find the tuning of a model, one should in principle consider
the tuning of all observables, since the absence of tuning in one
observable does not necessarily imply it is small in others,
see e.g. \cite{Schuster:2005py}.
In this paper, however, only the tuning of EWSB will be considered.

Finally, note that the FT with respect to a single parameter is by
definition (\ref{Eqn:FTab}) zero if that parameter happens to be zero
at the messenger scale.
An extreme version of this is found in the no-scale model \cite{Lahanas:1986uc},
where all scalar soft masses are
much smaller than the gaugino masses at the high scale.
Setting them to zero, and using (\ref{Eqn:FTab}) and (\ref{Eqn:FTdefn}) the FT could
be expected to be small.
However, it may be shown that this does not minimize the FT, since $M_3$ and
$\mu$ need to be quite large at the high scale to satisfy all the low-energy
experimental bounds (see \cite{BasteroGil:1999gu}).
In the results presented in this paper, no parameter is found to be zero
at the high scale.

\section{Minimal Model Independent Tuning}\label{Sec:MMIT}

In this section the minimal model independent tuning will be found as a
function of the messenger scale.

\subsection{Discussion of Minimization Procedure and Constraints}\label{Sec:MFTProcedure}

The FT given by equation (\ref{Eqn:FTdefn}) is written in terms of parameters
evaluated at the messenger scale.
In order to find the minimal FT (MFT) for a given messenger scale that is consistent
with low-energy experimental constraints, it is easiest to rewrite the FT expression in terms of
parameters that are evaluated at the low scale.
This can be done by expressing each high-scale parameter in terms of low-scale
parameters, see Appendix \ref{Sec:Master Formula}.
Once the FT is written in terms of low-scale parameters, $m_{H_u}^2(m_Z)$ may be
eliminated by using equation ($\ref{Eqn:EWSBmZ2}$) (neglecting contributions from
$m_{H_d}^2$).

The low-energy constraints considered in this paper include bounds on the (physical)
sparticle masses, on the gaugino masses, and on the Higgs mass\footnote{Constraints from
measurements of $\bsg$ or the electroweak $S$- and $T$-parameter do not significantly
affect the results presented below, since an experimentally consistent value
can be obtained by only small adjustments (if at all necessary) in the
least fine-tuned parameters - see also \cite{Essig:2007vq}.}.
The top quark mass $m_t$ is set to the central value of the latest
Tevatron mass measurement of 170.9 $\pm$ 1.8 GeV \cite{TevatronTop:2007bx}.
The physical stop masses are required to be at least 100 GeV which
is illustrative of the actual, slightly model dependent,
lower bound obtained from the Tevatron \cite{PDBook}.
It is found that the region of MFT does not quite saturate this bound,
although a slightly larger value for $m_t$ would allow the lighter stop
to be as low as 100 GeV.
The gaugino masses $M_1$ and $M_2$, as well as $\mu$, are
taken to have a lower bound of 100 GeV.
The gluino mass is found to be never smaller than 335 GeV
in the numerical results presented in this section, and this does not generically
violate any experimental bounds.

The most important constraint is the Higgs mass bound of
114.4 GeV (valid in the decoupling limit), since it turns out that
this bound is always saturated when minimizing the FT.
In the numerical results presented in this paper, the Higgs mass
is calculated using the formulas found in \cite{Sparticles:2004}
(see also \cite{Okada:1990vk,Ellis:1990nz,Haber:1990aw,Haber:1996fp,Carena:1995bx,Carena:2002es}).
These formulas include the one-loop corrections coming from the top/stop sector and are
simple enough to be used as constraints in the FT minimization (but note that the sign convention
used for $A_t$ is that of \cite{Martin:1997ns}).
A running top mass $m_t(m_t)$, evaluated in the $\overline{MS}$-scheme, is used to
capture some of the leading two-loop contributions.
Higher-order corrections to the Higgs mass still play
a very important role, however, and more accurate Higgs masses may be
obtained with the program $\mathtt{FeynHiggs}$.
In order to take some of these higher-order corrections into account and thus obtain a more
accurate estimate of the MFT, a lower bound for the Higgs mass
of 121.5 GeV is used in the FT minimization, instead of the SM lower bound of 114.4 GeV.
The typical low energy sparticle spectrum obtained in the analysis then
leads to Higgs masses that satisfy the SM Higgs bound when calculated with $\mathtt{FeynHiggs}$
(version 2.6.0, assuming real parameters).

Sequential Quadratic Programming (SQP) is used as a minimization
algorithm.  Given the FT function (\ref{Eqn:FTdefn}) written in terms of low scale parameters,
as well as linear constraints on the gaugino masses and $\mu$, non-linear constraints on
the physical stop and Higgs masses, and an initial guess, SQP %computes a search direction and
generates a less FT point until the minimum is found.
Unlike other minimization algorithms, SQP can handle arbitrary
constraints which is essential here due to the highly non-linear physical
stop mass and Higgs mass constraints.

\subsection{Numerical Results}\label{Sec:NumRes}

\begin{figure}[t]\begin{center}\includegraphics[scale=0.5]{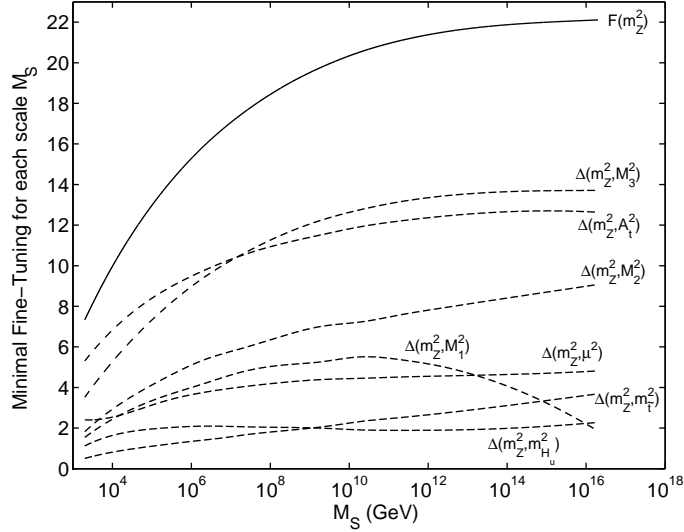}
\caption{The minimal fine-tuning as a function of the messenger scale $M_S$
for $\tan\beta=10$.  The top black line is the total minimal fine-tuning as defined
in equation (\ref{Eqn:FTdefn}) which includes all the individual contributions.
The individual contributions to the fine-tuning from $\mu^2$, $m_{H_u}^2$,
the gaugino masses $M_1^2$, $M_2^2$ and $M_3^2$, and the stop soft trilinear
coupling $A_t^2$ are included.  Moreover, the average fine-tuning of the stop
soft masses $m^2_{\tilde{t}_L}$ and $m^2_{\tilde{t}_R}$ is included as
in equation (\ref{Eqn:AvStopTuning}).
\label{Fig:MFT_vs_MS}}
\end{center}\end{figure}

Figure \ref{Fig:MFT_vs_MS} shows a plot of the MFT as a function of the
messenger scale $M_S$.
Shown are the individual contributions $\Delta\big(m_Z^2,m_i^2(M_S)\big)$ to the
FT, with $m_i^2$ given by $M_3^2$, $M_2^2$, $M_1^2$, $A_t^2$, $\mu^2$, or $m_{H_u}^2$.
The FT of $m_{\tilde{t}_L}^2$ and $m_{\tilde{t}_R}^2$ have been included as

\beq\label{Eqn:AvStopTuning}
\Delta(m_Z^2,m^2_{\tilde{t}}) = \left({\frac{1}{2}\Big[\left(\Delta(m_Z^2,m^2_{\tilde{t}_L})\right)^2+\left(\Delta(m_Z^2,m^2_{\tilde{t}_R})\right)^2\Big]}\right)^{1/2}.
\eeq
The (top) black line shows the total FT as defined by (\ref{Eqn:FTdefn}).

From the plot it is clear that the MFT increases as a function of the
messenger scale $M_S$.
This is expected since a higher messenger scale implies more RG running to the
low scale so that small differences in high-scale input parameters are magnified.
For $M_S=M_{\mbox{\scriptsize{GUT}}}$, the total MFT is about 22, i.e. 4.5$\%$.
(As an aside, for $\tan\beta=30$ and $m_A=1000$, the MFT for a Higgs mass of
114 GeV is about 11, i.e. 9$\%$.)
The largest contribution to the total minimal FT comes from
$M_3^2$ and $A_t^2$ which are both comparable for all values of $M_S$.
The next most important contribution is that from $M_2^2$.
The contributions from $\mu^2$, as well as $m_{\tilde{t}_L}^2$ and
$m_{\tilde{t}_R}^2$ are less important and increase only slightly as a function of $M_S$.
The FT from $m_{H_u}^2$ is very small for all messenger scales while the
contribution from $M_1^2$ is negligible for small and large $M_S$ but larger for intermediate messenger scales.

\begin{figure}[t]\begin{center}\includegraphics[scale=0.5]{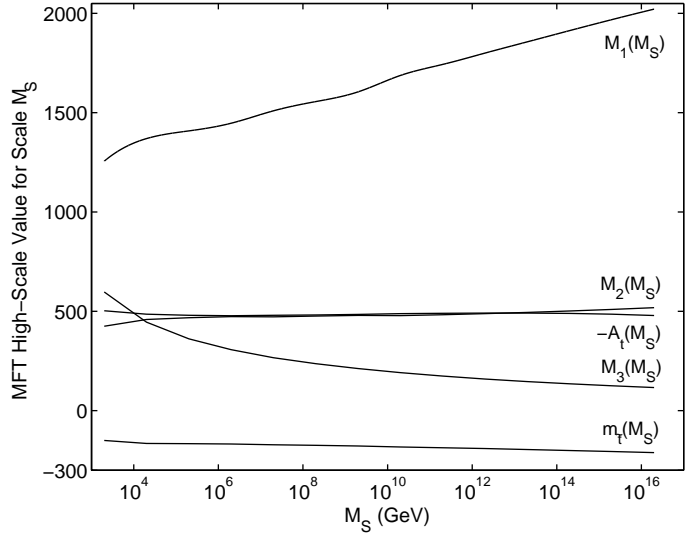}
\caption{The messenger scale values of $M_3$, $M_2$, $M_1$, $A_t$ and the average of the
stop soft masses squared, $m_{\tilde{t}}$, that give the minimal
fine-tuning (MFT) as a function of the messenger scale $M_S$ and for $\tan\beta=10$.
The high-scale values of $M_2$ and $A_t$, and to a lesser extent $M_1$ and $m_{\tilde{t}}$,
in the minimal fine-tuned region are roughly constant.
The high-scale value of $M_3$, however, decreases significantly as the messenger scale is
increased.
The reason for this is that the coefficient of $M_3^2$ in the expression for $m_Z^2$ increases
as a function of $M_S$, and thus the minimal fine-tuned region requires the value of $M_3$ to decrease as
$M_S$ increases.
\label{Fig:HighScaleM1M2M3At_vs_MS}}
\end{center}\end{figure}

The large contribution from $M_3^2$ is mainly because it has the largest (in magnitude) coefficient in
the expression for $m_Z^2$, at least for $M_S$ $\gtrsim$ $10^{10}$ GeV, see Figure \ref{Fig:coefficients}.
The coefficients of the cross-terms $A_t M_3$, $M_2 M_3$ and $M_1 M_3$ are smaller (see Appendix \ref{Sec:FTComps}), but together
still contribute about 40$\%$ of the FT with respect to $M_3^2$ for $M_S = M_{\mbox{\scriptsize{GUT}}}$.
The reason that the cross-term contributions are so large is that the MFT values of $A_t$,
$M_2$, and $M_1$ are rather sizeable at the messenger scale when compared with $M_3$
(at least for $M_S$ $\gtrsim$ $10^4$ GeV).
This is depicted in Figure \ref{Fig:HighScaleM1M2M3At_vs_MS}.

The FT of $m_Z^2$ with respect to $A_t^2$ is also very large even though the coefficients of $A_t^2$
and the cross-terms $A_t M_3$, $A_t M_2$ and $A_t M_1$ in the expression for $m_Z^2$ are rather small
(for $M_S=M_{\mbox{\scriptsize{GUT}}}$, about 50$\%$ of the FT comes from the cross-terms).
This is again because $A_t$, $M_2$ and $M_1$ are sizeable at $M_S$.
The contribution to the FT from $M_2^2$ is large for similar reasons.

The FT with respect to $\mu^2$ increases only slightly as a function of $M_S$ since the
coefficient of $\mu^2$ in the expression for $m_Z^2$ does not vary much, and
since the high-scale value of $\mu^2$ increases only slightly as $M_S$ is increased.
The contribution from $\mu^2$ is smaller than those from $M_3^2$, $M_2^2$ and $A_t^2$ because
the value of $\mu$ is comparatively small and also because there are no cross-terms in
the FT expression that involve $\mu$ and other (large) soft parameters.
Similar reasoning holds for the contributions from $m_{H_u}^2$, $m_{\tilde{t}_L}^2$ and
$m_{\tilde{t}_R}^2$.

\begin{table}\begin{center}\begin{tabular}{|c|c|c|c|}
\hline $A_t$ & $\sqrt{\frac{1}{2}(m_{\tilde{t}_L}^2+\,m_{\tilde{t}_R}^2})$ & $m_{\tilde{t}_1}$ & $m_{\tilde{t}_2}$ \\
\hline -610 GeV & 305 GeV & 110 GeV %113 GeV more precisely
& 475 GeV \\
\hline
\end{tabular}\caption{Low-scale values for the stop soft trilinear coupling, the average of the left- and
right-handed stop soft masses and the two physical stop masses.
These low scale values give the minimal fine-tuning for arbitrary messenger scales.
\label{Table:low scale stop sector}}\end{center}\end{table}

\begin{figure}[t]\begin{center}\includegraphics[scale=0.5]{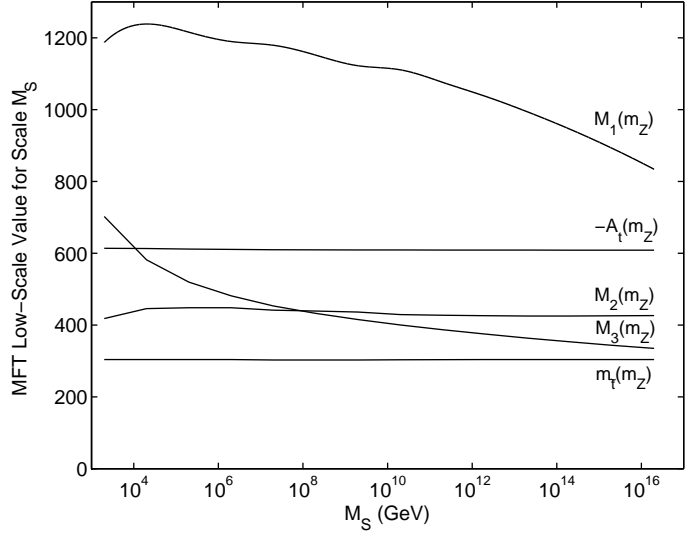}
\caption{The low-scale values of the gaugino masses $M_1$, $M_2$ and $M_3$, the
stop soft trilinear coupling $A_t$ and the average of the stop soft masses squared
$m_{\tilde{t}}$ that give the minimal
fine-tuning (MFT) for the messenger scale $M_S$ (with $\tan\beta=10$).
While the low-scale values of $M_2$, $A_t$ and $m_{\tilde{t}}$ that give the minimal
fine-tuning are roughly the same for all $M_S$,
the values of $M_1$ and $M_3$ decrease for larger $M_S$.
\label{Fig:LowScaleM1M2M3At_vs_MS}}
\end{center}\end{figure}

The low-energy spectrum that gives the MFT for a given messenger scale
remains roughly unchanged as the messenger scale changes.
The value of the stop soft trilinear coupling at the low scale is always about -610 GeV, with the
two physical stop masses around 110 GeV and 475 GeV, respectively,
see Table \ref{Table:low scale stop sector} and Figure \ref{Fig:LowScaleM1M2M3At_vs_MS}.
These values of the stop-sector parameters are essentially
determined by the constraint on the Higgs mass and from the
minimization of $\Delta(m_Z^2,m_{H_u}^2(M_S))$.
The ratio $X_t/m_{\tilde{t}}$ is approximately -2, where $X_t \equiv A_t - \mu\cot\beta$,
and $m_{\tilde{t}}$ $\equiv$ $\sqrt{\frac{1}{2}(m_{\tilde{t}_L}^2+\,m_{\tilde{t}_R}^2})$.
The MFT is thus found for the \emph{natural maximal-mixing} scenario which approximately
maximizes the radiative corrections to the Higgs sector for a given set of parameters and for \emph{negative}
$A_t$ \cite{Essig:2007vq,Heinemeyer:1999be,Kane:2004tk,Kitano:2006gv}.
Small deviations of $A_t$ (and to a lesser extent $m_{\tilde{t}_L}$ and $m_{\tilde{t}_R}$)
from its MFT value at the low scale lead to a very large
increase in the FT, mainly from $\Delta(m_Z^2,m_{H_u}^2(M_S))$.
This can be seen from (\ref{Eqn:mHusqM}), which shows that the
largest coefficients in the expression for $m_{H_u}^2(M)$ in
terms of low-scale parameters all involve powers of $A_t$.
Note that for generic points in the still allowed parameter space,
$\Delta(m_Z^2,m_{H_u}^2)$ would give one of the largest contribution to the FT.
To minimize the FT it is thus best to minimize $\Delta(m_Z^2,m_{H_u}^2(M_S))$
which essentially determines the values of the stop-sector parameters (see the
discussion in Section \ref{Sec:Explanation of NumResults}).
The other contributions to the FT are then not at their minimum, but they
are much smaller and less sensitive to variations in the parameters.

The low-scale values of the gaugino masses that give the MFT
for a given messenger scale are shown in Figure \ref{Fig:LowScaleM1M2M3At_vs_MS}.
While the value of $M_2$ that gives the MFT is roughly the same
for all $M_S$, the values of $M_1$ and $M_3$ decrease for larger $M_S$.
Changing $M_1$ away from its MFT value does affect the FT but
not excessively so, while a change in $M_3$ has a larger effect.
The $\mu$-parameter is always found to be less than 150 GeV for the MFT region
at any messenger scale.
Choosing it to be closer to 100 GeV instead has a negligible impact on the FT, and
allows a neutralino to be the lightest SM superpartner (LSP), instead of the lighter stop,
which is found to be the LSP in the numerical minimization procedure.

\begin{figure}[t]\begin{center}\includegraphics[scale=0.5]{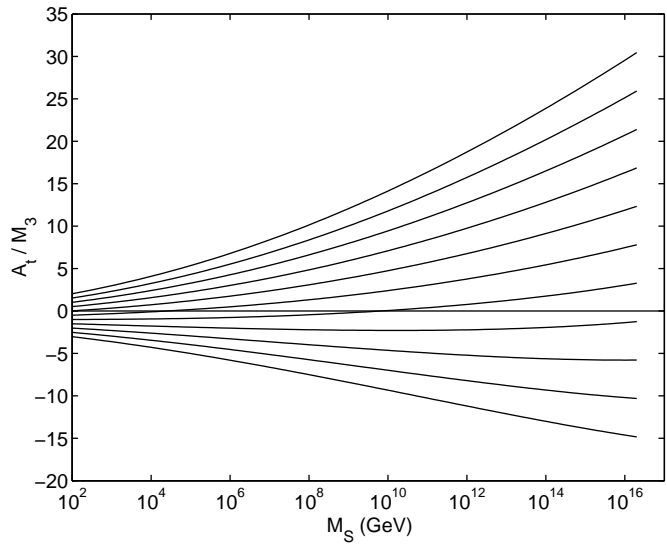}
\caption{The RG-evolution of $A_t/M_3$ for various low-scale boundary conditions
$A_t(m_Z)/M_3(m_Z)$ = $\{-2.0, -1.5, \ldots, 1.5, 2.0\}$ and $\tan\beta=10$.
The strongly attractive infrared quasi-fixed point near $A_t/M_3 \simeq -1$ is clearly visible.
The gaugino masses have been set to their minimal fine-tuned
values for the case $M_S=M_{\mbox{\scriptsize{GUT}}}$, i.e. $M_3(m_Z) \simeq 335$ GeV,
$M_2(m_Z) \simeq 430$ GeV, and $M_1(m_Z) \simeq 830$ GeV.
\label{Fig:At_fixed_pt_vs_MS}}
\end{center}\end{figure}

Negative $A_t$ may be expected to lead to less FT than positive $A_t$
because $A_t$ has a strongly attractive infrared quasi-fixed point near
\cite{Ferreira:1995sn,Lanzagorta:1995ai}
\beq\label{Eqn:Atfp}
A_t \simeq -M_3.
\eeq
(This relation is strictly valid only at the Pendleton-Ross quasi-fixed point
for the top Yukawa \cite{Pendleton:1980as}, and neglecting $SU(2)_L$ and $U(1)_Y$
gauge interactions.)
Because of this it is most natural for $A_t$
and $M_3$ to have opposite sign and be comparable in magnitude at
low scales due to renormalization group evolution, see Figure \ref{Fig:At_fixed_pt_vs_MS}.
For positive $A_t$ and maximal-mixing in the stop-sector, $A_t$ would have to be an order of
magnitude larger then $M_3$ at the messenger scale (see Figure \ref{Fig:At_fixed_pt_vs_MS})
which would lead to a much more FT parameter region.
The MFT region here does not satisfy (\ref{Eqn:Atfp}) exactly, but
instead $A_t/M_3 \simeq -1.8$ at the low scale, for $M_S=M_{\mbox{\scriptsize{GUT}}}$.
In order to satisfy (\ref{Eqn:Atfp}) exactly, $M_3$ would have
to be larger (assuming $A_t$ remains fixed).
This would increase the size of the stop masses under RG evolution as can be seen
from their $\beta$-functions, see (\ref{Eqn:BetastopL}) and (\ref{Eqn:BetastopR}), which would
lead to increased FT.

\begin{figure}[t]\begin{center}\includegraphics[scale=0.5]{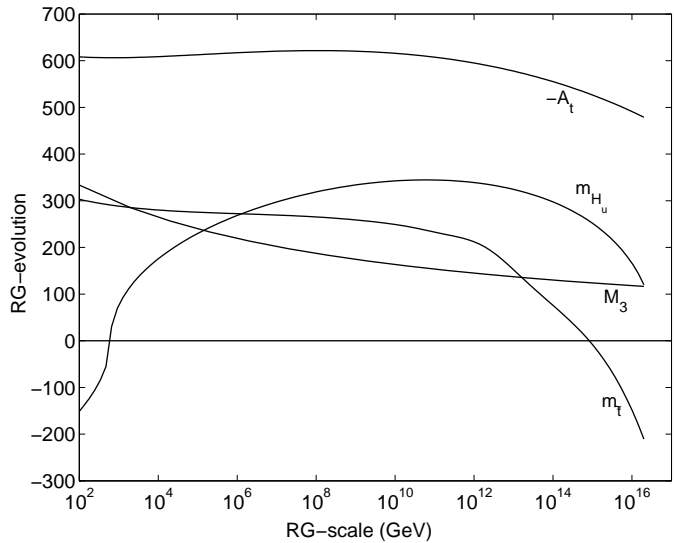}
\caption{The RG-trajectories of the minimal fine-tuned region
if the messenger scale is $M_S$=$\MGUT$ ($\tan\beta$ has been set to 10).
At the scale $m_Z$, the parameter values are $m_{\tilde{t}}\simeq$ 305 GeV,
$m_{\tilde{t}_1}$ $\simeq$ 110 GeV, $m_{\tilde{t}_2}$ $\simeq$ 475 GeV,
$M_3(m_Z) \simeq 335$ GeV, and $\mu(m_Z)=140$ GeV.
The minimal fine-tuned value is obtained for \emph{natural maximal-mixing}, i.e.
$A_t\simeq -2\MSusy$.
\label{Fig:RGevolution}}
\end{center}\end{figure}

The MTMSSM has negative soft squark squared masses at the messenger
scale (see also \cite{Dermisek:2006ey}).
This remains the case even if the messenger scale is very low and only on the
order of a few TeV (for very low messenger scales, finite threshold corrections
should really be included).
Under RG-evolution the masses get driven positive very quickly within about
a decade of running.
It is the sizeable values of the gaugino masses that pull them up towards positive
values.
For smaller messenger scales the MFT region has a larger gluino mass,
which drives the squark masses to positive values even faster while running
towards the infrared.
Equations (\ref{Eqn:AtM}) and (\ref{Eqn:mstLsqM}) or (\ref{Eqn:mstRsqM})
in Appendix \ref{Sec:Master Formula} show that
negative squarks at the messenger scale lead to more stop-mixing at the low scale,
as was pointed out in \cite{Dermisek:2006ey}.
Figure \ref{Fig:RGevolution} shows the RG-trajectories of the MFT region
if the messenger scale is $M_S$=$\MGUT$.

The presence of tachyonic squarks at the messenger scale
\cite{Frere:1983ag,Derendinger:1983bz} and/or
very large $A_t$ \cite{Gunion:1987qv,Casas:1995pd} may lead to
dangerous color and/or charge breaking (CCB) minima.

Very large $A_t$ may result in dangerous CCB minima around the
EW scale.
These CCB minima occur in the $(\tilde{t}_L, \tilde{t}_R, H_u)$ plane \cite{LeMouel:2001ym}.
The condition that the EW minimum is the global minimum may be estimated by
going along the D-flat direction $|\tilde{t}_L|$ = $|\tilde{t}_R|$ = $|H_u|$ and is given by \cite{Kusenko:1996jn}
\begin{equation}
A_t^2+3\mu^2\lsim3(m_{\tilde{t}_L}^2+m_{\tilde{t}_R}^2).
\end{equation}
Assuming instead that the EW minimum is only metastable but has a large enough lifetime
gives the weaker constraint \cite{Kusenko:1996jn}
\begin{equation}
A_t^2+3\mu^2\lsim7.5(m_{\tilde{t}_L}^2+m_{\tilde{t}_R}^2).
\end{equation}
The MTMSSM easily satisfies the second condition, as well as satisfying the first condition.
There are thus no dangerous CCB minima resulting from large $A_t$.

Tachyonic stops at the messenger scale may result in an unbounded
from below potential along D-flat directions involving the stop fields,
as well as first and/or second generation squark fields or slepton fields.
Loop corrections give rise to an effective potential which is not
unbounded from below, but they generically introduce a CCB minimum
with a vacuum expectation value (VEV) on the order of the messenger scale.
The MTMSSM may thus have CCB minima with a VEV around the EW
scale if the messenger scale is low, or CCB minima with a VEV large compared
to the EW scale if the messenger scale is high.
Since the EW minimum is metastable and long-lived for
$m_{\tilde{t}}\gsim\frac{1}{6}M_3$ \cite{Riotto:1995am}, it turns out that
these CCB minima are not dangerous in the MTMSSM.
Moreover, the MTMSSM does not determine the masses of the sleptons
or first and second generation squarks since these do not play an important role in
the FT.
It is thus always possible to choose them in such a way to avoid CCB
minima without changing the above FT results.

Finally, it is interesting to note that there are several near degenerate parameter subspaces
along which the FT does not change much.
The first and second generation particles and their superpartners do not contribute much to
the FT because in equation (\ref{Eqn:mZsq generic}) they appear only with a small coefficient.
The parameter $S_Y$ is also not very important for the same reason.
A more interesting near degenerate subspace is that the FT is rather insensitive
to changes in the \emph{difference} of the two stop soft mass squared parameters
at the low scale as long as their sum is kept fixed.
This may be understood from the expression for $m_Z^2$, e.g. equation (\ref{Eqn:MF1}),
in which only their sum appears (using the one-loop RG equations).
However, even with only one-loop RG equations this degeneracy is not exact
since small discrepancies appear in the FT measure from
equations (\ref{Eqn:mstLsqM}) and (\ref{Eqn:mstRsqM}).  Moreover, the difference in the two
stop soft mass squared parameters appears in the calculation of the physical stop masses and
this affects the size of the Higgs mass, which is the most crucial low-energy
constraint when calculating the FT.
The FT only starts to change by an order one number when
$\sqrt{|m_{\tilde{t}_L}^2-m_{\tilde{t}_R}^2|}\sim300$ GeV for $M_S=\MGUT$.

\subsection{Analytic Motivation for Numerical Results}\label{Sec:Explanation of NumResults}

The numerical results presented in section \ref{Sec:NumRes}
may be motivated analytically.
The discussion will for now assume $M_S=\MGUT$,
but generalizes to arbitrary $M_S$ with a few caveats discussed below.

In order to get a physical Higgs mass satisfying the experimental bound
without generating large FT for the EWSB, it is natural to maximize the
radiative corrections to $m_h$.
Due to the strongly attractive quasi-fixed point for $A_t$, this is achieved for
negative $A_t$ near (natural) maximal mixing (at least for $m_h$ not too large, see
Section \ref{Sec:FT_vs_mh}).

The most important contribution to the FT comes from $\Delta(m_Z^2,m_{H_u}^2(M_S))$
since it has the largest coefficients, see Appendix \ref{Sec:FTComps}.
Eliminating $\hat{m}_{H_u}^2$ with the EWSB equation (\ref{Eqn:MF1}) and using
the average stop soft mass squared
$\hat{m}_{\tilde{t}}^2 = (\hat{m}_{\tilde{t}_L}^2 + \hat{m}_{\tilde{t}_R}^2)/2$
gives
\begin{eqnarray}\label{Eqn:FTmHu}
m_Z^2\Delta(m_Z^2,\hat{m}_{H_u}^2) &=& |-m_{Z}^2 -\,2.19\, \hat{\mu}^2
+\, 1.36\, \hat{m}_{\tilde{t}}^2 +\, 5.24\, \hat{M}_3^2 \nonumber \\
      &   &  -\, 0.44\, \hat{M}_2^2 +\, 0.46\, \hat{M}_3\, \hat{M}_2 -\,
0.77\, \hat{A}_t\,\hat{M}_3 -\, 0.17\, \hat{A}_t\, \hat{M}_2 \nonumber \\
      &   &  -\, 0.01 \hat{M}_1^2 +\, 0.22\, \hat{A}_t^2|.
\end{eqnarray}
It is possible to have cancelations among the various terms in this expression.
$\Delta(m_Z^2,M_3^2(M_S))$ also has large coefficients,
but cancelations among its terms are impossible since $\hat{A}_t$ is negative
(see Appendix \ref{Sec:FTComps}).

Ignoring $\hat{\mu}^2$, cancelation of the largest
terms in equation (\ref{Eqn:FTmHu}), i.e. the gluino term and the
average stop soft mass squared term,
decreases the FT by setting  $\hat{m}_{H_u}^2\simeq m_{H_u}^2$ and leads to
tachyonic squarks at the messenger scale \cite{Dermisek:2006ey}
\begin{equation}
\hat{m}_{\tilde{t}}^2\simeq-3.9\hat{M}_3^2.
\end{equation}
Next, the four terms on the second line of equation (\ref{Eqn:FTmHu}) can
cancel by taking
\begin{equation}
\hat{M}_3\simeq\frac{0.96\hat{M}_2+0.37\hat{A}_t}{1-1.67\frac{\hat{A}_t}{\hat{M}_2}}.
\end{equation}
Assuming $\hat{M}_2\simeq-\hat{A}_t$, this simplifies to $\hat{M}_2\simeq 4.5\hat{M}_3$.
Furthermore, keeping only the most important terms, the natural maximal-mixing
scenario implies
\begin{eqnarray}
-2\simeq\frac{A_t}{m_{\tilde{t}}} &\simeq&
(0.32\hat{A_t}-2.13\hat{M}_3-0.27\hat{M}_2-0.03\hat{M}_1)\left[0.66\hat{m}_{\tilde{t}}^2+5.15\hat{M}_3^2\right.\nonumber\\
 &&
\left.+0.11\hat{M}_2^2+0.02\hat{M}_1^2+0.19\hat{A}_t\hat{M}_3+0.04\hat{A}_t\hat{M}_2-0.05\hat{A}_t^2\right]^{-1/2}\nonumber\\
 &=&
(-4.80\hat{M}_3-0.03\hat{M}_1)\left[2.16\hat{M}_3^2+0.02\hat{M}_1^2\right]^{-1/2}
\end{eqnarray}
which leads to $\hat{M}_1 \simeq 15\hat{M}_3$, again assuming $\hat{M}_2\simeq -\hat{A}_t$.
It is now possible to compute the ratio of the soft trilinear coupling with the
gluino mass at the EWSB scale,
\begin{equation}
\frac{A_t}{M_3}\simeq\frac{0.32\hat{A_t}-2.13\hat{M}_3-0.27\hat{M}_2-0.03\hat{M}_1}{2.88\hat{M}_3}\simeq-1.8.
\end{equation}
These results agree well with the numerical results presented in section \ref{Sec:NumRes}.

Note that a GUT scale model which predicts degenerate and negative squark
and slepton soft masses at the GUT scale would need very large wino and bino masses in
comparison to the gluino mass in order to drive the slepton soft masses
to positive values under RG running to the EWSB scale \cite{Dermisek:2006qj}.
This is due to the small coefficients of the bino and wino masses in the
$\beta$-functions of the slepton soft masses.
It is interesting that the MFT region prefers the bino mass larger than the
wino mass and, in turn, the wino mass larger than the gluino mass.

Although this cancelation pattern holds to a good approximation for higher
messenger scales, $\hat{m}_{\tilde{t}}^2$ does not exactly cancel
$\hat{M}_3^2$ as the messenger scale decreases.
For lower messenger scales, $\hat{m}_{\tilde{t}}^2$ becomes less
tachyonic while $\hat{M}_3^2$ increases, allowing the stop
masses to be driven positive faster under RG running to the EWSB scale.
Moreover, the coefficient of $\hat{M}_3^2$ in the expression for $m_Z^2$
(\ref{Eqn:mZsq generic}) decreases
significantly, as can be seen in Figure \ref{Fig:coefficients}.
Therefore the cancelation pattern in $\Delta(m_Z^2,\hat{m}_{H_u}^2)$
discussed above does not hold since the $\hat{m}_{\tilde{t}}^2$
contribution decreases while the $\hat{M}_3^2$ term gives
a comparable contribution for all messenger scales
(except for very small messenger scales).
On the other hand, being a supersymmetric parameter, $\hat{\mu}$ and its
coefficient in equation (\ref{Eqn:mZsq generic}) does not change much for
different messenger scales.
Compared to $\hat{M}_3^2$ and $\hat{m}_{\tilde{t}}^2$, its contribution
becomes important at lower messenger scales and a lower FT can be obtained
by canceling the three contributions together.
The other relations in the above cancelation pattern holds to a good
approximation for lower messenger scales, although for $M_S\lsim10^5$
the cancelation pattern becomes more involved.

\subsection{Summary of Phenomenological Implications}

The above analysis shows that the MTMSSM has small values for $\mu$,
the stop masses and the gluino mass.
The gluino in the MTMSSM is around 335 GeV for $M_S=\MGUT$, but heavier for lower $M_S$.
There is large mixing in the stop-sector which introduces a significant splitting
between the two physical stop masses.
They have masses of around 115 GeV and 475 GeV respectively, see
Table \ref{Table:low scale stop sector}.
Thus the MTMSSM may have a stop as the LSP.
However, as mentioned before, $\mu$ can be chosen to be small enough so that a
neutralino is the LSP without affecting FT by much.

At the Large Hadron Collider, gluino pair-production in the
MTMSSM is thus rather large and comparable to top quark pair-production.
The production of $\tilde{t}_1\tilde{t}_1$ is also of the same order.

The gluinos are Majorana particles, and can decay into the lightest stop via
$\tilde{g}\tilde{g} \to tt\tilde{t}_1\tilde{t}_1$ producing same-sign top
quarks 50$\%$ of the time.
The top quarks each decay into $Wb$, and the events with two same-sign
top quarks will contain two same-sign leptons if the $W$ decays leptonically.
If a neutralino and a chargino are lighter than the stop, the decay
$\tilde{t}_1 \to \chi_1^+ b$ is possible, with $\chi_1^+$ further decaying
into a neutralino and soft jets or leptons.
The events thus also contain missing energy and a number of $b$-jets, some of
which are soft if the $\tilde{t}_1 - \chi^+_1$ mass splitting is small.

If $\tilde{t}_1$ is the LSP a number of further interesting signatures
are possible, see \cite{Culbertson:2000am}.
The lighter stop can either be pair-produced directly or from gluino decays.
Even though it is the lightest SM superpartner, it may decay into a lighter
goldstino $\tilde{G}$ via the flavor-violating decay $\tilde{t}_1 \to c \tilde{G}$
or via the three-body decay $\tilde{t}_1 \to b W \tilde{G}$.
The decay rate depends on the messenger scale, with lower messenger scales leading to
larger decay rates.
For reasonable messenger scales, its decay length easily exceeds the hadronization
length scale, and the stop in general hadronizes before it decays \cite{Culbertson:2000am}.
For messenger scales less than a few hundred TeV, the decay length is small enough so that
the decay products seem to originate from the interaction region.
The three-body decay leads to a similar signature as the top decay but can be
distinguished from it, see \cite{Chou:1999zb}.
For larger messenger scales, $\tilde{t}_1$ decays inside a hadronized mesino or sbaryon
and a variety of interesting signatures are possible \cite{Culbertson:2000am},
including mesino-anti-mesino oscillations \cite{Sarid:1999zx}.

Another interesting possibility is the direct pair-production of the heavier stop $\tilde{t}_2$.
Since the two physical stop masses are split by a large amount, the decay mode
$\tilde{t}_2 \to \tilde{t}_1 + Z$ is kinematically allowed and has a
sizeable branching ratio \cite{Perelstein:2007nx}.
The resulting signature depends on the $\tilde{t}_1$ decay channel as discussed above.
For $\chi_1^+$ and $\chi^1_0$ lighter than $\tilde{t}_1$, the authors of
\cite{Perelstein:2007nx} propose to look for the inclusive
signature $Z(l^+,l^-) bb E_T\!\!\!\!\!\!\!/ ~~ X$, where the two leptons $l^+$ and
$l^-$ have an invariant mass equal to the $Z$-mass.
Detecting this signature would give evidence for the maximal-mixing scenario but
requires a large integrated luminosity (at least $\mathcal{O}$(100 fb$^{-1}$))
\cite{Perelstein:2007nx}.
Since the mass difference between $\tilde{t}_1$ and the LSP is small in the MTMSSM
this signature will be very hard to see since the jet from the decay $\tilde{t}_1 \to \chi_1^+ b$
is soft which makes it more difficult to separate the signal from the SM background \cite{Perelstein:2007nx}.

An alternative way to measure the parameters in the stop-sector is to use
the Higgs boson as a probe \cite{Dermisek:2007fi}.
A measurement of the Higgs mass and its production rate in the gluon
fusion channel allows the average of the two stop soft masses as well as the stop mixing
to be determined in many regions of the still allowed MSSM parameter space, and especially
in regions where the FT is small \cite{Dermisek:2007fi}.

\subsection{Fine-Tuning with Respect to Other Parameters}

This subsection briefly discusses other parameters that may in principle
contribute to the FT.

If the goal is to find the MFT region of a model and make a prediction
of what parameter region is preferred for the model from a FT point of view,
there is no reason to include the FT of experimentally known parameters
such as $g_Y$, $g_2$ ,$g_3$, or $\lambda_t$.
Taking into account the known parameters in the minimization procedure
would most likely lead to other MFT values for all parameters,
including MFT values for the known parameters which would
in all likelihood not match the experimental values.

If the goal, however, is to find the FT of a given model, one should
in principle include contributions from experimentally known parameters.
For example, FT with respect to $\lambda_t$, $\Delta(m_Z^2,\lambda_t(M_S))$,
may give a large contribution to the total FT due to the large top mass.
Indeed, with the MFT values for $M_S=M_{\mbox{\scriptsize{GUT}}}$,
$\Delta(m_Z^2,\lambda_t(M_{\MGUT}))$ $\simeq$ 8.
This, however, increases the total FT only by a small
amount from 22.1 to 23.5.

What about FT with respect to $m_{12}^2$ and $\tan\beta$?
These parameters are unknown and in principle they should be
included in the minimization procedure.
With the help of equation (\ref{Eqn:EWSBmZ1}) and symmetries,
it is however easy to see that $\Delta(m_Z^2,m_{12}^2(M_S))=0$.
Indeed $m_{12}^2$ does not appear directly in the expression for $m_Z^2$.
Furthermore it breaks a $U(1)_{\mbox{\scriptsize{PQ}}}$- and a $U(1)_R$-symmetry
and consequently does not feed back into any other $\beta$-functions
since no other parameter breaks both symmetries.
Thus $m_{12}^2$ cannot appear in equation (\ref{Eqn:EWSBmZ1}) and is
therefore completely free, which allows $m_A$ to be chosen accordingly
as discussed in Section \ref{Sec:EWSB}.

The FT of $\tan\beta$ has not been taken into account in the
minimization procedure since an explicit expression for
$m_Z^2$ can only be obtained assuming a specific value
for $\tan\beta$, because $\lambda_t$ depends on
$\tan\beta$ through $m_t$.
Moreover, since $\tan\beta$ is then a free parameter the
approximation leading to equation (\ref{Eqn:EWSBmZ2}) may not be valid
anymore and $m_{H_d}^2$ should be reintroduced.
Contributions from bottom/sbottom and tau/stau sectors should also be included if
$\tan\beta$ becomes large.

\section{Minimal Fine-Tuning as a Function of the Higgs Mass} \label{Sec:FT_vs_mh}

\begin{figure}[t]\begin{center}\includegraphics[scale=0.5]{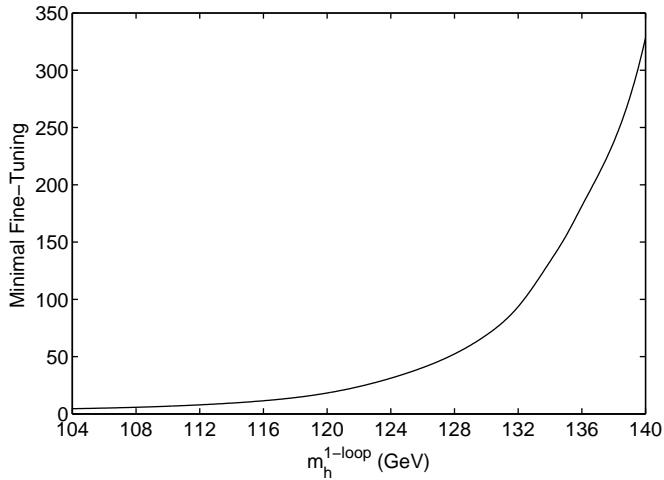}
\caption{The minimal fine-tuning as a function of the lower bound on the
Higgs mass $m_h$, where the calculation of $m_h$ only includes the one-loop
corrections from the top-stop sector ($\tan\beta=10$, $m_A=250$ GeV, $m_t=170.9$ GeV).
\label{Fig:FT_vs_mh_oneloop}}
\end{center}\end{figure}

The Higgs mass $m_h$ is the most important low-energy constraint that determines the
amount of minimal fine-tuning (MFT).
It is therefore interesting to look at how the MFT is affected when the lower bound
on $m_h$ is changed.
Figure \ref{Fig:FT_vs_mh_oneloop} shows a plot of the MFT as a function of the lower bound
on $m_h$, where the calculation of $m_h$ is the same one used in the FT minimization described
in Section \ref{Sec:MFTProcedure}, and only includes the one-loop
corrections from the top-stop sector (with $m_A=250$ GeV, $\tan\beta=10$, $m_t=170.9$ GeV, and $M_S=\MGUT$).
The Higgs mass calculated with the one-loop corrections will be denoted by $m_h^{1\ell}$.
The region of MFT always saturates the bound on $m_h^{1\ell}$ and has negative $A_t$.
The minimal FT is about 1$\%$ for $m_h^{1\ell}\simeq 132$ GeV.

There are, however, other important one-loop and two-loop corrections that can
significantly affect $m_h$, and these need to be included in order to get a more
accurate idea of how the MFT changes as a function of the lower bound on $m_h$.
With these additional corrections, $m_h$ is not anymore a symmetric function of the stop-mixing parameter
$X_t = A_t-\mu\cot\beta \simeq A_t$, where the latter approximation is good for sizeable $\tan\beta$.
It can be up to 5 GeV larger for $X_t=+2 \MSusy$ than for $X_t = -2
\MSusy$, the difference arising from non-logarithmic two-loop
contributions to $m_h$, see \cite{Espinosa:1999zm,Carena:2000dp,Heinemeyer:2004ms}.
Moreover, large chargino masses, i.e. large values of $M_2$ and $\mu$, can give
important negative contributions to $m_h$ \cite{Carena:1999xa}.
These corrections are also not included in $m_h^{1\ell}$.
Two-loop corrections that allow the gluino mass to affect $m_h$ can also
be important but are smaller in general - this will be ignored in the following
discussion since the impact on the results presented below is negligible.

\begin{figure}\begin{center}\includegraphics[scale=0.5]{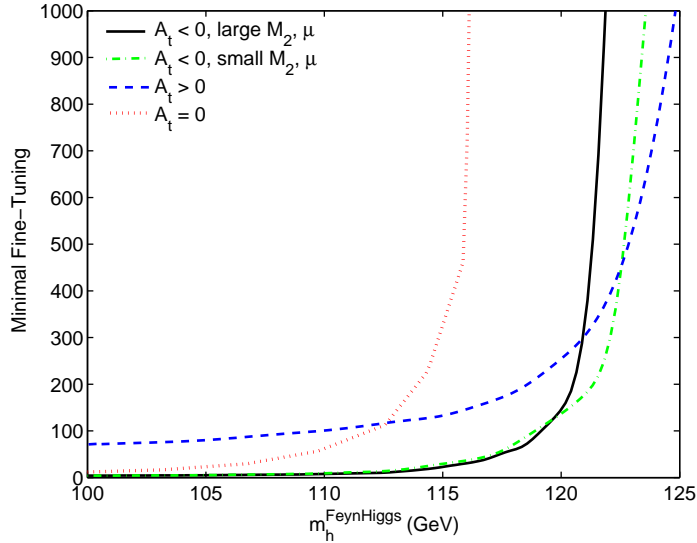}
\caption{The minimal fine-tuning as a function of the lower bound on the
Higgs mass $m_h$ calculated with $\mathtt{FeynHiggs}$ 2.6.0
($\tan\beta=10$, $m_A=250$ GeV, $m_t=170.9$ GeV).
Throughout this paper the fine-tuning is minimized subject to a constraint on
$m_h$, where $m_h$ is estimated with a one-loop formula as described in
Section \ref{Sec:MFTProcedure}.
The different lines arise from different assumptions made about $A_t$, or
$\mu$ and $M_2$, when minimizing the fine-tuning.
These different assumptions give rise to different low-energy spectra that
present the least fine-tuned parameter choices satisfying these assumptions.
These low-energy spectra may then be used in $\mathtt{FeynHiggs}$ to
calculate $m_h$.
Although $M_2$, $\mu$ and the sign of $A_t$ do not affect the one-loop estimate of
$m_h$ which only contains the dominant corrections, they do
affect the $\mathtt{FeynHiggs}$ estimate of $m_h$.
For the solid black line no constraint was set on $A_t$, and $\mu$ and
$M_2$ were only required to be above 100 GeV.
It is the same line as in Figure \ref{Fig:FT_vs_mh_oneloop}, but with
$m_h$ estimated by $\mathtt{FeynHiggs}$ instead of the one-loop
formula.
The dashed blue line assumes $A_t$ is positive and near maximal mixing,
also with $M_2$ and $\mu$ only required to be above 100 GeV.
The dash-dot green curve makes no assumption about $A_t$ but sets $\mu=100$ GeV and
$M_2=100$ GeV.
The dotted red line assumes $A_t=0$, and again only requires $\mu$ and $M_2$ to be
larger than 100 GeV.
Further details and explanations are given in the text.
\label{Fig:FT_vs_mh}}
\end{center}\end{figure}

The MFT spectrum that was found with the minimization procedure may be
used to calculate $m_h$ with $\mathtt{FeynHiggs}$.
The $\mathtt{FeynHiggs}$ estimate for $m_h$ will be denoted by $m_h^{\mathtt{FH}}$.
The result is the solid black line in Figure \ref{Fig:FT_vs_mh}.
This MFT spectrum characteristically has large chargino masses and
a negative value for $A_t$ near the ``natural'' maximal mixing scenario.

Comparing the solid black line in Figure \ref{Fig:FT_vs_mh} with the curve in
Figure \ref{Fig:FT_vs_mh_oneloop} shows the well-known fact that the higher-order
corrections to $m_h$ are extremely important.
There are two additional very striking features.
First of all, as $m_h^{\mathtt{FH}}$ increases and approaches 120 GeV, the FT increases enormously.
Any further small increase in the Higgs mass results in an exponentially large increase in the FT.
The reason is that as $m_h^{\mathtt{FH}}$ approaches 120 GeV here, it only grows
logarithmically as a function of the stop masses.
The stop masses therefore become exponentially large and thus increase the FT
exponentially (see also \cite{Essig:2007vq}).

The second striking feature of this curve is that the value of the Higgs mass at which the
FT starts to increase exponentially is rather low (the MFT is already $1\%$ for
$m_h^{\mathtt{FH}}\simeq 119$ GeV).
This value of $m_h$ may be increased by just under 2 GeV by choosing larger $\tan\beta$ and $m_A$
(recall that throughout this discussion $\tan\beta=10$ and $m_A=250$ GeV).
Note that the latest Tevatron top mass value ($m_t$ = 170.9 GeV) has been used in the calculation,
and a slightly different value can also change $m_h$ by a few GeV.

An obvious question is whether the MFT region is significantly different if
$m_h^{\mathtt{FH}}$ were used in the minimization procedure instead of $m_h^{1\ell}$ (the
former is too complicated to be used).
For MSSM spectra that give small $m_h$ this is certainly not the case, since there is not a very
large discrepancy between the two Higgs mass estimates $m_h^{1\ell}$ and $m_h^{\mathtt{FH}}$.
The difference between the two Higgs mass estimates becomes significant, however, for
MSSM spectra that give a large $m_h$, and the approximation $m_h^{\mathtt{FH}}$ can be substantially
smaller than $m_h^{1\ell}$.
Also, as mentioned above, $m_h^{\mathtt{FH}}$ can be substantially larger for positive $A_t$ (near maximal mixing) than for
negative $A_t$ (near ``natural'' maximal mixing), and increases as the chargino masses decrease.
On the other hand, $m_h^{1\ell}$ remains unaffected by the sign of $A_t$ and the size of the chargino masses.
It is thus possible that the MFT region does not
coincide with the region obtained in the above minimization procedure
as the lower bound on $m_h$ increases.
This is indeed the case, as will now be discussed.

The FT may be minimized with the constraint that the chargino masses are small.
Since the effect of varying $\mu$ and $M_2$ on the FT are noticeable but not
substantial, the resulting spectrum will be characterized by gluino and stop masses that
are only slightly larger than those obtained in the MFT region discussed in this paper.
The value of $A_t$ is still negative.
This spectrum may be used to calculate $m_h^{\mathtt{FH}}$.
The result is shown by the dash-dot green curve in Figure \ref{Fig:FT_vs_mh}.
For $m_h^{\mathtt{FH}}$ not too large, the solid black curve lies below the dash-dot green
curve because the MFT region has large values of $M_2$, see Section \ref{Sec:MMIT}.
As $m_h^{\mathtt{FH}}$ increases further, however, the FT becomes exponentially large since
the stop masses become exponentially large.
Smaller chargino masses lead to larger values of $m_h^{\mathtt{FH}}$, and the two curves show
that for $m_h$ just below 120 GeV, a smaller FT may be obtained by decreasing the size of $M_2$.
This behavior cannot be captured by $m_h^{1\ell}$ which is unaffected by a change in the
chargino masses.
Note that the transition between the two regions described by the two curves is smooth, and that
it occurs when the MFT is already more than 1$\%$.

Next, the FT may be minimized with the constraint that $A_t$ is positive and near maximal
mixing.
The resulting low-energy spectrum is characterized by small chargino and gluino masses.
This spectrum may then be used to calculate $m_h^{\mathtt{FH}}$, and the
MFT as a function of this value of $m_h^{\mathtt{FH}}$ is displayed
by the dashed blue line in Figure \ref{Fig:FT_vs_mh}.
Comparing the solid black line or dash-dot green line with the dashed blue line, it is clear
that for small $m_h^{\mathtt{FH}}$ the MFT region has negative values of $A_t$.
Even though negative $A_t$ might be expected to always give less FT than positive $A_t$ due
to the IR quasi-fixed point, the increase in $m_h^{\mathtt{FH}}$ by several GeV
by making $A_t$ positive is substantial, and as $m_h^{\mathtt{FH}}$ approaches about 123 GeV,
the two curves cross.
Thus, there is a transition from $A_t$ $\simeq$ $-2\MSusy$ to $A_t$ $\simeq$ $+2\MSusy$ of
the minimal fine-tuned region as $m_h^{\mathtt{FH}}$ increases.
This behavior is again not captured by $m_h^{1\ell}$ which is independent of the sign
of $A_t$.
The transition occurs when the minimal FT is already quite large (about 0.2$\%$).

This transition from negative to positive $A_t$ is not smooth, in the sense that the
first derivative of the curve at the transition point is not continuous\footnote{One may perhaps
refer to this as the \emph{first order phase transition of fine-tuning}.}.
To show this, the FT may be minimized with the constraint $A_t=0$.
The resulting low-energy spectrum may then again be used to calculate $m_h^{\mathtt{FH}}$,
and the result is shown by the dotted red line in Figure \ref{Fig:FT_vs_mh}.
The value of $m_h^{\mathtt{FH}}$ for vanishing stop-mixing, $A_t=0$, is much lower than for
the two maximal mixing scenarios, $A_t \simeq \pm2\MSusy$, and it is clear that the
MFT region does not interpolate smoothly between them as a function of $A_t$.

The main point of the analysis in this section is that although the MSSM is already fine-tuned
at least at about the $5\%$ level (if the messenger scale equals the GUT scale), there is not much
room left for the Higgs mass to increase before the FT becomes exponentially worse.

Note that for a lower messenger scale the Higgs mass can have a slightly larger value
before the MFT begins to increase enormously.
For example, for $M_S$ = 200 TeV, the MFT is 1$\%$ for $m_h\simeq 123$ GeV.
So even for a lower messenger scale the Higgs mass cannot be that much beyond 120 GeV before the
MFT increases dramatically.

\section{Conclusions}\label{Sec:Conclusion}

This paper presented the minimally tuned Minimal Supersymmetric Standard Model (MTMSSM).
The MSSM parameter region that has the
minimal model-independent fine-tuning (FT) of EWSB was found.
Model - independent means that no relations were assumed between the soft SUSY
breaking parameters at the scale at which they are generated, here
referred to as the messenger scale.
Instead, all of the important parameters were allowed to be
independent and free at the messenger scale, and were taken to
contribute to the total FT of the EWSB scale.
The messenger scale itself was varied between 2 TeV and $\MGUT$
and the effect of this on the minimal FT was presented.

The most important parameters that contribute to the tuning are $|\mu|^2$, $m_{H_u}^2$,
the gaugino masses $M_1$, $M_2$ and $M_3$, the stop soft masses $m^2_{\tilde{t}_L}$
and $m^2_{\tilde{t}_R}$, and the stop soft trilinear coupling $A_t$.
The MSSM spectra which lead to the minimal
model-independent FT were found by numerically minimizing the
FT expression subject to constraints on the Higgs, stop, and gaugino
masses (the Higgs mass was found to always be the most important low-energy constraint).
The high-energy spectra are characterized by tachyonic stop soft masses, even for messenger
scales as low as 2 TeV (but note that threshold effects in the RG-running were neglected
throughout).
The potential existence of charge and/or color breaking minima turns out not to
be a problem.
The gluino mass, $M_3$, is much smaller than the wino mass, $M_2$,
and $M_2$ in turn is much smaller than the bino mass $M_1$.
The low-scale spectra are characterized by negative $A_t$ near the maximal
mixing scenario that maximizes the Higgs mass.
The large stop mixing leads to a large splitting between the two stop
mass eigenstates.
Interesting phenomenological signatures include the possibility of a stop LSP.

The minimal FT was also found as a function of the lower bound on the Higgs mass (with
the messenger scale set to $\MGUT$).
Although in the numerical minimization procedure the dominant one-loop
expression for $m_h$ was used as a constraint, the resulting least fine-tuned spectra
were used to calculate $m_h$ more accurately with $\mathtt{FeynHiggs}$.
A plot of the minimal FT as a function of $m_h$ was presented.
There are several striking features of this plot.
For $m_h$ larger than about 120 GeV the FT increases very rapidly.
This value of $m_h$ is rather low, perhaps surprisingly so.
It is only slightly dependent on the parameters in the Higgs sector.
Near it, the value of $A_t$ in the least FT region also makes a sudden
transition from lying near $-2\MSusy$ to lying near $+2\MSusy$, where $\MSusy$ is the
average of the two stop soft masses.
The upshot of this particular analysis is that although the MSSM is already
fine-tuned at least at about the $5\%$ level (if the messenger scale equals the GUT scale),
there is not much room left for the Higgs mass to increase before the FT becomes
exponentially worse.

\newpage

\noindent {{\bf \large Acknowledgements}}

\vspace{0.1cm}

We thank S.~Thomas for suggesting this problem and for
many enlightening and useful discussions.
We also thank R.~Derm\'{\i}\v{s}ek for helpful discussions,
T.~Banks for useful suggestions and S.~Heinemeyer for answering
questions about $\mathtt{FeynHiggs}$.
This research is supported by the Department of Physics
and Astronomy at Rutgers University.  JFF is also
supported by the FQRNT.

\vspace{-0.1cm}

\appendix

\section{Semi-numerical Solutions of the MSSM \\ One-Loop RG-Equations}\label{Sec:Master Formula}

This appendix reviews the procedure for solving the
MSSM one-loop RG equations semi-numerically \cite{Ibanez:1983di,Carena:1996km}.
The low scale $M_0$ is set to be $m_Z$, and the high (messenger) scale $M_S$ is taken
to lie anywhere between $m_Z$ and $M_{\mbox{\scriptsize{GUT}}}$.
Threshold corrections are neglected when solving the RG-equations.

The main goal is to obtain an expression for $m_Z^2$ in terms of high-scale input
parameters as in equation (\ref{Eqn:mZsq generic}).
Assuming that $\tan\beta$ is not too small, this requires solving $|\mu(m_Z)|^2$
and $m_{H_u}^2\!(m_Z)$ in terms of high-scale parameters (for moderate values of
$\tan\beta$, $m_{H_d}^2$ may be neglected, see equation (\ref{Eqn:EWSBmZ2})).
The fine-tuning may then be calculated and naturally expressed in terms of high-scale
parameters as in equation (\ref{Eqn:FTdefn}).
However, in order to minimize the fine-tuning taking into account low-scale constraints
on the Higgs, stop, and gaugino masses, it is more appropriate to rewrite the fine-tuning expression
in terms of low scale parameters.
This requires that $\mu$ as well as all the soft supersymmetry breaking parameters appearing in equation
(\ref{Eqn:FTdefn}) be written in terms of low scale parameters.

In solving the RG-equations, only the contributions from the third generation
particles will be included, since the third generation Yukawa couplings are much
larger than those from the first and second generations.
Moreover, the contributions from the bottom/sbottom and tau/stau sectors are
neglected as $\tan\beta$ is taken to be not too large.

The high-scale parameters may in general be written in terms of low scale-parameters as
\begin{equation}
m_i^2(M_S)=\sum_{j,k}\,c_{ijk}(\tan\beta, M_0, M_S)\,m_j(M_0)\,m_k(M_0).
\end{equation}
For example, for $M_S=M_{\mbox{\scriptsize{GUT}}}$, the
expressions for the most important high-scale parameters written in
terms of low-scale parameters are
\begin{eqnarray}
\hat{M}_i & = & d_i\,M_i \;\;\;\;\;\;\;\;\;\;\;\;\;\;\;\;\;\;\;\;\;\{d_1,d_2,d_3\}=\{2.42,1.22,0.35\}\label{Eqn:MiM}\\
\hat{A}_t & = & 3.15\, A_t +\, 2.33\, M_3 +\, 1.03\, M_2 +\, 0.26\, M_1\label{Eqn:AtM}\\
\hat{m}_{H_u}^2 & = & 2.07\, m_{H_u}^2 +\, 1.07\, m_{\tilde{t}_L}^2 +\, 1.07\, m_{\tilde{t}_R}^2 +\, 0.19\, M_3^2 -\, 0.98\, M_2^2\nonumber\\
                                       &   & -\, 0.31\, M_1^2 +\, 3.38\, A_t^2 +\, 3.69\, A_t\,M_3 +\, 1.19\, A_t\,M_2 +\, 0.24\, A_t\,M_1\nonumber\\
                                       &   & +\, 0.76\, M_3\,M_2 +\, 0.15\, M_3\,M_1 +\, 0.05\, M_2\,M_1 +\, 0.06\, S_Y\label{Eqn:mHusqM}\\
\hat{m}_{\tilde{t}_L}^2 & = & 0.36\, m_{H_u}^2 +\, 1.36\, m_{\tilde{t}_L}^2 +\, 0.36\, m_{\tilde{t}_R}^2 -\, 0.72\, M_3^2 -\, 0.81\, M_2^2\nonumber\\
                                       &   & -\, 0.06\, M_1^2 +\, 1.13\, A_t^2 +\, 1.23\, A_t\,M_3 +\, 0.40\, A_t\,M_2 +\, 0.08\, A_t\,M_1\nonumber\\
                                       &   & +\, 0.25\, M_3\,M_2 +\, 0.05\, M_3\,M_1 +\, 0.02\, M_2\,M_1 +\, 0.02\, S_Y\label{Eqn:mstLsqM}\\
\hat{m}_{\tilde{t}_R}^2 & = & 0.72\, m_{H_u}^2 +\, 0.72\, m_{\tilde{t}_L}^2 +\, 1.72\, m_{\tilde{t}_R}^2 -\, 0.65\, M_3^2 -\, 0.18\, M_2^2\nonumber\\
                                       &   & -\, 0.46\, M_1^2 +\, 2.26\, A_t^2 +\, 2.46\, A_t\,M_3 +\, 0.80\, A_t\,M_2 +\, 0.16\, A_t\,M_1\nonumber\\
                                       &   & +\, 0.50\, M_3\,M_2 +\, 0.10\, M_3\,M_1 +\, 0.04\, M_2\,M_1 -\, 0.09\, S_Y\label{Eqn:mstRsqM}\\
\hat{\mu} & = & 0.95\, \mu. \label{Eqn:muM}
\end{eqnarray}
Similar type of expressions hold for low-scale parameters as a function of high-scale parameters.
The gauge couplings $g_\alpha$, $\alpha\in\{1,2,3\}$, and the top Yukawa coupling $\lambda_t$
are fixed at the low scale by their experimental values \cite{PDBook}.
Section \ref{Sec:A1} gives the solution of their RG-equations.

The MSSM one-loop $\beta$-functions that need to be solved come in three different functional forms \cite{Martin:1993zk}.
The RG-equations of the gaugino masses $M_\alpha$, the supersymmetric Higgsino mass $\mu$, and $S_Y$ are of the form
\begin{equation}
\frac{dm_i}{dt}=f_i(\lambda_t,g_\alpha)\,m_i,\;\;\;m_i\in\{M_\alpha,\mu,S_Y\},
\end{equation}
where $t=\ln(M_S/M_0)$.  Their solution is given by
\begin{equation}\label{Eqn:group1sol}
m_i(t)=m_i(0)\,\exp{\int_0^tdt'\,f_i(\lambda_t,g_\alpha)}.
\end{equation}

The stop soft trilinear coupling has the functional form
\begin{equation}
\frac{dA_t}{dt}=a(\lambda_t)\,A_t+b(g_\alpha,M_\alpha).
\end{equation}
The solution of this equation is more involved due to the presence of both homogeneous and inhomogeneous terms,
and requires the solution for the gaugino masses (\ref{Eqn:group1sol}).  It may be written as (see Section \ref{Sec:Agroup2})
\begin{equation}\label{Eqn:group2sol}
A_t(t)=e^{\int dt'a(\lambda_t)}\,A_t(0)+e^{\int dt'a(\lambda_t)}\int_0^tdt'e^{-\int dt''a(\lambda_t)}\,b(g_\alpha,M_\alpha).
\end{equation}

Finally, the RG-equations of the up-type Higgs soft mass and the stop soft masses form a system of coupled inhomogeneous differential equations,
\begin{equation}\label{Eqn:group3}
\frac{dm_i^2}{dt}=\sum_ju_{ij}(\lambda_t)m_j^2+v_i(g_\alpha,M_\alpha,S_Y,A_t),\;\;\;m_i^2\in\{m_{H_u}^2,m_{\tilde{t}_L}^2,m_{\tilde{t}_R}^2\}.
\end{equation}
This may be solved (see Section \ref{Sec:Agroup3}) using the solutions for the gaugino masses and $S_Y$ (\ref{Eqn:group1sol}) as well as the solution for the
stop soft trilinear coupling (\ref{Eqn:group2sol}),
\begin{equation}\label{Eqn:group3sol}
m_i^2(t)=\left(e^{\int dt'u(\lambda_t)}\,m^2(0)+e^{\int dt'u(\lambda_t)}\int_0^tdt'e^{-\int dt''u(\lambda_t)}\,v(g_\alpha,M_\alpha,S_Y,A_t)\right)_i.
\end{equation}

\subsection{Gauge and Yukawa Couplings}\label{Sec:A1}

The one-loop $\beta$-functions for the gauge and top Yukawa couplings in the MSSM are
\begin{eqnarray}
8\pi^2\beta_{g_\alpha^2} &=& b_\alpha \,g_\alpha^4,\;\;\;\{b_Y,b_2,b_3\}=\{11,1,-3\}\\
16\pi^2\beta_{\lambda_t} &=& \lambda_t\,\left(6\,\lambda_t^2-\frac{16}{3}\,g_3^2-3\,g_2^2-\frac{13}{9}\,g_Y^2\right).
\end{eqnarray}
Their solutions are
\begin{eqnarray}
g_\alpha^2(t) &=& g_\alpha^2(0)\,\xi_\alpha^{-1}(t)\\
\lambda_t^2(t) &=& \lambda_t^2(0)\,E(t;\vec{n}_0)\,G(t;\vec{n}_0)^{-1},\label{Eqn:soltop}
\end{eqnarray}
where $\vec{n}_0=\left(\frac{13}{9b_1},\frac{3}{b_2},\frac{16}{3b_3}\right)=\left(\frac{13}{99},3,-\frac{16}{9}\right)$, and for future convenience the functions
\begin{eqnarray}
\xi_\alpha(t) &=& 1-\frac{b_\alpha}{8\pi^2}\,g_\alpha^2(0)t\\
E(t;\vec{n}) &=& \prod_{\alpha=1}^3\,\xi_\alpha^{(\vec{n})_\alpha}(t)\\
F(t;\vec{n}) &=& \int_0^t\,dt'\,E(t';\vec{n})\\
G(t;\vec{n}) &=& 1-\frac{3}{4\pi^2}\,\lambda_t^2(0)\,F(t;\vec{n})
\end{eqnarray}
have been introduced.
The solution (\ref{Eqn:soltop}) is analytic if $g_2$ and $g_Y$ are set to zero \cite{Hill:1980sq,Lanzagorta:1995gp},
whereas non-zero values of $g_2$ and $g_Y$ require a numerical integration.

\subsection{Gaugino Masses, $\mu$-term and $S_Y$}\label{Sec:Agroup1}

The RG-equations for the gaugino masses, $\mu$ and $S_Y$ are
\begin{eqnarray}
\beta_{M_\alpha} &=& \frac{M_\alpha}{g_\alpha^2}\,\beta_{g_\alpha^2}\\
16\pi^2\beta_\mu &=& \mu\,\left(3\,\lambda_t^2-3\,g_2^2-g_Y^2\right)\\
8\pi^2\beta_{S_Y} &=& g_Y^2\sum_{\mbox{{\tiny scalars i}}}\left(\frac{Y_i}{2}\right)^2S_Y.
\end{eqnarray}
The general solution is of the form (\ref{Eqn:group1sol}), and may be written as
\begin{eqnarray}
M_\alpha(t) &=& M_\alpha(0)\,\xi_\alpha^{-1}(t)\label{Eqn:gaugino}\\
\mu(t) &=& \mu(0)\,G(t;\vec{n}_0)^{-\frac{1}{4}}\,\xi_2^{\frac{3}{2}}(t)\,\xi_1^{\frac{1}{22}}(t)\label{Eqn:mu}\\
S_Y(t) &=& S_Y(0)\,\xi_1^{-1}(t)
\end{eqnarray}
with the notation of Section \ref{Sec:A1}.  The solutions for the
gaugino masses and $S_Y$ are analytic while $\mu$ must be solved numerically
unless the contributions from $g_2$ and $g_Y$ are neglected.

\subsection{Stop Soft Trilinear Coupling}\label{Sec:Agroup2}

The $\beta$-function of the stop soft trilinear coupling is
\begin{equation}
8\pi^2\beta_{A_t}=\left(6\,\lambda_t^2\,A_t-\frac{16}{3}\,g_3^2\,M_3-3\,g_2^2\,M_2-\frac{13}{9}\,g_Y^2\,M_1\right).
\end{equation}
Using the solutions for the gaugino masses (\ref{Eqn:gaugino}), this equation may be integrated and written as
\begin{equation}\label{Eqn:At}
A_t(t)=\frac{1}{G(t;\vec{n}_0)}\left[A_t(0)+\sum_{\alpha=1}^3(\vec{n}_0)_\alpha\frac{M_\alpha(0)}{\xi_\alpha(t)}\Big(G(t;\vec{n}_0)-\xi_\alpha(t)\,G(t;\vec{n}_0-\vec{e}^\alpha)\Big)\right]
\end{equation}
where $(\vec{e}^\alpha)_\beta=\delta_\beta^\alpha$ are the usual unit vectors.  If $g_2$ and $g_Y$ are zero, the solution does not
require a numerical integration.

\subsection{Up-type Higgs Soft Mass and Stop Soft Masses}\label{Sec:Agroup3}

The $\beta$-functions of $m_{H_u}^2$, $m_{\tilde{t}_L}^2$ and $m_{\tilde{t}_R}^2$ are
\begin{eqnarray}
8\pi^2\beta_{m_{H_u}^2} &=& 3\lambda_t^2\left[m_{H_u}^2+m_{\tilde{t}_L}^2+m_{\tilde{t}_R}^2+|A_t|^2\right]  \nonumber \\
                            & &  \hspace{2cm} -3\,g_2^2\,|M_2|^2-g_Y^2\,|M_1|^2-\frac{1}{2}\,g_Y^2\,S_Y\\
8\pi^2\beta_{m_{\tilde{t}_L}^2} &=& \lambda_t^2\left[m_{H_u}^2+m_{\tilde{t}_L}^2+m_{\tilde{t}_R}^2+|A_t|^2\right] \nonumber \\
                                    & & \hspace{0.5cm}-\frac{16}{3}\,g_3^2\,|M_3|^2-3\,g_2^2\,|M_2|^2-\frac{1}{9}\,g_Y^2\,|M_1|^2-\frac{1}{6}\,g_Y^2\,S_Y\label{Eqn:BetastopL}\\
8\pi^2\beta_{m_{\tilde{t}_R}^2} &=& 2\lambda_t^2\left[m_{H_u}^2+m_{\tilde{t}_L}^2+m_{\tilde{t}_R}^2+|A_t|^2\right] \nonumber \\
                                    & & \hspace{2cm} -\frac{16}{3}\,g_3^2\,|M_3|^2-\frac{16}{9}\,g_Y^2\,|M_1|^2-\frac{2}{3}\,g_Y^2\,S_Y.\label{Eqn:BetastopR}
\end{eqnarray}
They form a system of coupled inhomogeneous differential equations.
Note that $A_t$ appears quadratically in these $\beta$-functions which gives
cross-terms between $M_\alpha(0)$ and $A_t(0)$ (see equation (\ref{Eqn:At})).
The equations can be solved as in (\ref{Eqn:group3sol}) but it is possible to simplify
the analysis by the change of variables
\begin{eqnarray}
X &=& m_{H_u}^2-m_{\tilde{t}_L}^2-m_{\tilde{t}_R}^2\\
Y &=& m_{H_u}^2-3m_{\tilde{t}_L}^2\\
Z &=& m_{H_u}^2+m_{\tilde{t}_L}^2+m_{\tilde{t}_R}^2.
\end{eqnarray}
In terms of the new variables, the $\beta$-functions are
\begin{eqnarray}
8\pi^2\beta_X &=& \frac{32}{3}\,g_3^2\,|M_3|^2+\frac{8}{9}\,g_Y^2\,|M_1|^2+g_Y^2\,S_Y\\
8\pi^2\beta_Y &=& 16\,g_3^2\,|M_3|^2+6\,g_2^2\,|M_2|^2-\frac{2}{3}\,g_Y^2\,|M_1|^2\\
8\pi^2\beta_Z &=& 6\lambda_t^2Z+6\lambda_t^2|A_t|^2-\frac{32}{3}\,g_3^2\,|M_3|^2-6\,g_2^2\,|M_2|^2-\frac{26}{9}g_Y^2|M_1|^2.
\end{eqnarray}
In this form, $\beta_X$ and $\beta_Y$ are easily integrated since they have no homogeneous term
(which is due to the fact that the corresponding matrix $u_{ij}$ in (\ref{Eqn:group3}) has $\mbox{rank}=1$)
\begin{eqnarray}
X(t) &=& X(0)-\frac{16}{9}\,M_3^2(0)\,\left(\xi_3^{-2}(t)-1\right)\\
     & & \hspace{2cm} +\frac{4}{99}\,M_1^2(0)\,\left(\xi_1^{-2}(t)-1\right)+\frac{1}{11}\,S_Y(0)\,\left(\xi_1^{-1}(t)-1\right)\nonumber\\
Y(t) &=& Y(0)-\frac{8}{3}\,M_3^2(0)\,\left(\xi_3^{-2}(t)-1\right)\\
     & & \hspace{2cm} +3\,M_2^2(0)\,\left(\xi_2^{-2}(t)-1\right)-\frac{1}{33}\,M_1^2(0)\,\left(\xi_1^{-2}(t)-1\right).\nonumber
\end{eqnarray}
The equation for $Z$ requires a numerical integration (even if $g_2$ and $g_Y$ are zero)
\begin{eqnarray}
Z(t) &=& \frac{1}{G(t;\vec{n}_0)}\left[Z(0)-\sum_{\alpha=1}^3(\vec{n}_0)_\alpha\frac{M^2_\alpha(0)}{\xi_\alpha^2(t)}\Big(G(t;\vec{n}_0)-\xi_\alpha^2(t)\,G(t;\vec{n}_0-2\vec{e}^\alpha)\Big)\right.\nonumber\\
 && \left.+\frac{3}{4\pi^2}\lambda_t^2(0)\int_0^tdt'\,E(t';\vec{n}_0)\,|A_t(t')|^2\right].
\end{eqnarray}
The solutions for $m_{H_u}^2$, $m_{\tilde{t}_L}^2$ and $m_{\tilde{t}_R}^2$ in terms of $X$, $Y$ and $Z$ are then
\begin{eqnarray}
m_{H_u}^2(t) &=& \frac{1}{2}\Big(X(t)+Z(t)\Big)\\
m_{\tilde{t}_L}^2(t) &=& \frac{1}{6}\Big(X(t)-2Y(t)+Z(t)\Big)\\
m_{\tilde{t}_R}^2(t) &=& \frac{1}{3}\Big(-2X(t)+Y(t)+Z(t)\Big).
\end{eqnarray}

\section{Fine-tuning Components}\label{Sec:FTComps}

This appendix lists for completeness the expressions for the fine-tuning of $m_Z^2$ with
respect to $M_3^2$, $M_2^2$, $M_1^2$, $\mu^2$, $A_t^2$, $m_{H_u}^2$, $m_{\tilde{t}_L}^2$ and $m_{\tilde{t}_R}^2$.
The fine-tuning components as a function of high-scale parameters are easily found
from the fine-tuning measure, equation (\ref{Eqn:FTab}), with the observable $m_Z^2$
written as in equation (\ref{Eqn:mZsq generic}).
For $M_S=\MGUT$, the fine-tuning components are
\begin{eqnarray}
m_Z^2\Delta(m_Z^2,\hat{M}_3^2) & \simeq & 5.24\hat{M}_3^2+0.23\hat{M}_3\hat{M}_2+0.03\hat{M}_3\hat{M}_1-0.38\hat{A}_t\hat{M}_3\\
m_Z^2\Delta(m_Z^2,\hat{M}_2^2) & \simeq &-0.44\hat{M}_2^2+0.23\hat{M}_3\hat{M}_2+0.01\hat{M}_2\hat{M}_1-0.08\hat{A}_t\hat{M}_2\\
m_Z^2\Delta(m_Z^2,\hat{M}_1^2) & \simeq &-0.01\hat{M}_1^2+0.03\hat{M}_3\hat{M}_1+0.01\hat{M}_2\hat{M}_1-0.01\hat{A}_t\hat{M}_1\\
m_Z^2\Delta(m_Z^2,\hat{\mu}^2) & \simeq & -2.19\hat{\mu}^2\\
m_Z^2\Delta(m_Z^2,\hat{A}_t^2) & \simeq &0.22\hat{A}_t^2-0.38\hat{A}_t\hat{M}_3-0.08\hat{A}_t\hat{M}_2-0.01\hat{A}_t\hat{M}_1\\
m_Z^2\Delta(m_Z^2,\hat{m}_{H_u}^2) & \simeq & -1.32\hat{m}_{H_u}^2\\
      & \simeq & -m_{Z}^2 -\,2.19\, \hat{\mu}^2 +\, 1.36\, \hat{m}_{\tilde{t}}^2 +\, 5.24\, \hat{M}_3^2 \nonumber \\
      &   &  -\, 0.44\, \hat{M}_2^2 +\, 0.46\, \hat{M}_3\, \hat{M}_2 -\,
0.77\, \hat{A}_t\,\hat{M}_3 -\, 0.17\, \hat{A}_t\, \hat{M}_2 \nonumber \\
      &   &  -\, 0.01 \hat{M}_1^2 +\, 0.22\, \hat{A}_t^2\nonumber\\
m_Z^2\Delta(m_Z^2,\hat{m}_{\tilde{t}_L}^2) & \simeq &0.68\hat{m}_{\tilde{t}_L}^2\\
m_Z^2\Delta(m_Z^2,\hat{m}_{\tilde{t}_R}^2) & \simeq &0.68\hat{m}_{\tilde{t}_R}^2.
\end{eqnarray}
Here it is understood that the absolute value of the right-hand sides of each of these equations
is meant to be taken.
The EWSB relation, equation (\ref{Eqn:MF1}), was used to eliminate $\hat{m}_{H_u}^2$.
It is natural to eliminate $\hat{m}_{H_u}^2$ instead of $\hat{\mu}^2$ or any other soft
supersymmetry breaking parameters since $\hat{\mu}^2$ is supersymmetric while the other soft
supersymmetry breaking parameters are not involved in the EWSB equation at the EW scale.
With the help of equations (\ref{Eqn:MiM})-(\ref{Eqn:muM}), it is now
straightforward to rewrite the FT expression (\ref{Eqn:FTdefn}) in terms of low-scale
parameters.

{\small{
\bibliographystyle{utcaps}
\bibliography{Bibliography}

\providecommand{\href}[2]{#2}\begingroup\raggedright\begin{thebibliography}{10}

\bibitem{ALEPH:2006cr}
{\bf ALEPH, DELPHI, L3 and OPAL} Collaboration, {T}he {L}{E}{P} {W}orking
  {G}roup {f}or {H}iggs~{B}oson {S}earches, ``Search for neutral {MSSM} {H}iggs
  bosons at {LEP},''
\href{http://arXiv.org/abs/hep-ex/0602042}{{\tt hep-ex/0602042}}.
%%CITATION = HEP-EX 0602042;%%.

\bibitem{Ellis:1986yg}
J.~R. Ellis, K.~Enqvist, D.~V. Nanopoulos, and F.~Zwirner, ``Observables in
  Low-Energy Superstring Models,'' {\em Mod. Phys. Lett.} {\bf A1} (1986)
57.
%%CITATION = MPLAE,A1,57;%%.

\bibitem{Barbieri:1987fn}
R.~Barbieri and G.~F. Giudice, ``Upper Bounds on Supersymmetric Particle
  Masses,'' {\em Nucl. Phys.} {\bf B306} (1988)
63.
%%CITATION = NUPHA,B306,63;%%.

\bibitem{deCarlos:1993yy}
B.~de~Carlos and J.~A. Casas, ``One loop analysis of the electroweak breaking
  in supersymmetric models and the fine tuning problem,'' {\em Phys. Lett.}
  {\bf B309} (1993) 320--328,
\href{http://arXiv.org/abs/hep-ph/9303291}{{\tt hep-ph/9303291}}.
%%CITATION = HEP-PH/9303291;%%.

\bibitem{deCarlos:1993ca}
B.~de~Carlos and J.~A. Casas, ``The Fine tuning problem of the electroweak
  symmetry breaking mechanism in minimal SUSY models,''
\href{http://arXiv.org/abs/hep-ph/9310232}{{\tt hep-ph/9310232}}.
%%CITATION = HEP-PH/9310232;%%.

\bibitem{Anderson:1994dz}
G.~W. Anderson and D.~J. Castano, ``Measures of fine tuning,'' {\em Phys.
  Lett.} {\bf B347} (1995) 300--308,
\href{http://arXiv.org/abs/hep-ph/9409419}{{\tt hep-ph/9409419}}.
%%CITATION = HEP-PH/9409419;%%.

\bibitem{Ciafaloni:1996zh}
P.~Ciafaloni and A.~Strumia, ``Naturalness upper bounds on gauge mediated soft
  terms,'' {\em Nucl. Phys.} {\bf B494} (1997) 41--53,
\href{http://arXiv.org/abs/hep-ph/9611204}{{\tt hep-ph/9611204}}.
%%CITATION = HEP-PH/9611204;%%.

\bibitem{Chankowski:1997zh}
P.~H. Chankowski, J.~R. Ellis, and S.~Pokorski, ``The fine-tuning price of
  LEP,'' {\em Phys. Lett.} {\bf B423} (1998) 327--336,
\href{http://arXiv.org/abs/hep-ph/9712234}{{\tt hep-ph/9712234}}.
%%CITATION = HEP-PH/9712234;%%.

\bibitem{Agashe:1997kn}
K.~Agashe and M.~Graesser, ``Improving the fine tuning in models of low energy
  gauge mediated supersymmetry breaking,'' {\em Nucl. Phys.} {\bf B507} (1997)
  3--34,
\href{http://arXiv.org/abs/hep-ph/9704206}{{\tt hep-ph/9704206}}.
%%CITATION = HEP-PH/9704206;%%.

\bibitem{Chan:1997bi}
K.~L. Chan, U.~Chattopadhyay, and P.~Nath, ``Naturalness, weak scale
  supersymmetry and the prospect for the observation of supersymmetry at the
  Tevatron and at the LHC,'' {\em Phys. Rev.} {\bf D58} (1998) 096004,
\href{http://arXiv.org/abs/hep-ph/9710473}{{\tt hep-ph/9710473}}.
%%CITATION = HEP-PH/9710473;%%.

\bibitem{Wright:1998mk}
D.~Wright, ``Naturally nonminimal supersymmetry,''
\href{http://arXiv.org/abs/hep-ph/9801449}{{\tt hep-ph/9801449}}.
%%CITATION = HEP-PH/9801449;%%.

\bibitem{Kane:1998im}
G.~L. Kane and S.~F. King, ``Naturalness implications of LEP results,'' {\em
  Phys. Lett.} {\bf B451} (1999) 113--122,
\href{http://arXiv.org/abs/hep-ph/9810374}{{\tt hep-ph/9810374}}.
%%CITATION = HEP-PH/9810374;%%.

\bibitem{BasteroGil:1999gu}
M.~Bastero-Gil, G.~L. Kane, and S.~F. King, ``Fine-tuning constraints on
  supergravity models,'' {\em Phys. Lett.} {\bf B474} (2000) 103--112,
\href{http://arXiv.org/abs/hep-ph/9910506}{{\tt hep-ph/9910506}}.
%%CITATION = HEP-PH/9910506;%%.

\bibitem{Casas:2003jx}
J.~A. Casas, J.~R. Espinosa, and I.~Hidalgo, ``The MSSM fine tuning problem: A
  way out,'' {\em JHEP} {\bf 01} (2004) 008,
\href{http://arXiv.org/abs/hep-ph/0310137}{{\tt hep-ph/0310137}}.
%%CITATION = HEP-PH/0310137;%%.

\bibitem{Casas:2004uu}
J.~A. Casas, J.~R. Espinosa, and I.~Hidalgo, ``A relief to the supersymmetric
  fine tuning problem,''
\href{http://arXiv.org/abs/hep-ph/0402017}{{\tt hep-ph/0402017}}.
%%CITATION = HEP-PH/0402017;%%.

\bibitem{Casas:2004gh}
J.~A. Casas, J.~R. Espinosa, and I.~Hidalgo, ``Implications for new physics
  from fine-tuning arguments. I: Application to SUSY and seesaw cases,'' {\em
  JHEP} {\bf 11} (2004) 057,
\href{http://arXiv.org/abs/hep-ph/0410298}{{\tt hep-ph/0410298}}.
%%CITATION = HEP-PH/0410298;%%.

\bibitem{Schuster:2005py}
P.~C. Schuster and N.~Toro, ``Persistent fine-tuning in supersymmetry and the
  NMSSM,''
\href{http://arXiv.org/abs/hep-ph/0512189}{{\tt hep-ph/0512189}}.
%%CITATION = HEP-PH/0512189;%%.

\bibitem{Dermisek:2005ar}
R.~Dermisek and J.~F. Gunion, ``Escaping the large fine tuning and little
  hierarchy problems in the next to minimal supersymmetric model and h $\to$ a
  a decays,'' {\em Phys. Rev. Lett.} {\bf 95} (2005) 041801,
\href{http://arXiv.org/abs/hep-ph/0502105}{{\tt hep-ph/0502105}}.
%%CITATION = HEP-PH/0502105;%%.

\bibitem{Dermisek:2006ey}
R.~Dermisek and H.~D. Kim, ``Radiatively generated maximal mixing scenario for
  the Higgs mass and the least fine tuned minimal supersymmetric standard
  model,'' {\em Phys. Rev. Lett.} {\bf 96} (2006) 211803,
\href{http://arXiv.org/abs/hep-ph/0601036}{{\tt hep-ph/0601036}}.
%%CITATION = HEP-PH 0601036;%%.

\bibitem{Casas:2006bd}
J.~A. Casas, J.~R. Espinosa, and I.~Hidalgo, ``Expectations for LHC from
  naturalness: Modified vs. SM Higgs sector,''
\href{http://arXiv.org/abs/hep-ph/0607279}{{\tt hep-ph/0607279}}.
%%CITATION = HEP-PH/0607279;%%.

\bibitem{Kobayashi:2006fh}
T.~Kobayashi, H.~Terao, and A.~Tsuchiya, ``Fine-tuning in gauge mediated
  supersymmetry breaking models and induced top Yukawa coupling,'' {\em Phys.
  Rev.} {\bf D74} (2006) 015002,
\href{http://arXiv.org/abs/hep-ph/0604091}{{\tt hep-ph/0604091}}.
%%CITATION = HEP-PH/0604091;%%.

\bibitem{Athron:2007ry}
P.~Athron and D.~J. Miller, ``A New Measure of Fine Tuning,''
\href{http://arXiv.org/abs/arXiv:0705.2241 [hep-ph]}{{\tt arXiv:0705.2241
  [hep-ph]}}.
%%CITATION = ARXIV:0705.2241;%%.

\bibitem{Frank:2006yh}
M.~Frank {\em et al.}, ``The Higgs boson masses and mixings of the complex MSSM
  in the Feynman-diagrammatic approach,''
\href{http://arXiv.org/abs/hep-ph/0611326}{{\tt hep-ph/0611326}}.
%%CITATION = HEP-PH 0611326;%%.

\bibitem{Degrassi:2002fi}
G.~Degrassi, S.~Heinemeyer, W.~Hollik, P.~Slavich, and G.~Weiglein, ``Towards
  high-precision predictions for the MSSM Higgs sector,'' {\em Eur. Phys. J.}
  {\bf C28} (2003) 133--143,
\href{http://arXiv.org/abs/hep-ph/0212020}{{\tt hep-ph/0212020}}.
%%CITATION = HEP-PH 0212020;%%.

\bibitem{Heinemeyer:1998np}
S.~Heinemeyer, W.~Hollik, and G.~Weiglein, ``The masses of the neutral CP-even
  Higgs bosons in the MSSM: Accurate analysis at the two-loop level,'' {\em
  Eur. Phys. J.} {\bf C9} (1999) 343--366,
\href{http://arXiv.org/abs/hep-ph/9812472}{{\tt hep-ph/9812472}}.
%%CITATION = HEP-PH 9812472;%%.

\bibitem{Heinemeyer:1998yj}
S.~Heinemeyer, W.~Hollik, and G.~Weiglein, ``FeynHiggs: A program for the
  calculation of the masses of the neutral CP-even Higgs bosons in the MSSM,''
  {\em Comput. Phys. Commun.} {\bf 124} (2000) 76--89,
\href{http://arXiv.org/abs/hep-ph/9812320}{{\tt hep-ph/9812320}}.
%%CITATION = HEP-PH 9812320;%%.

\bibitem{Heinemeyer:2007aq}
S.~Heinemeyer, W.~Hollik, H.~Rzehak, and G.~Weiglein, ``The Higgs sector of the
  complex MSSM at two-loop order: QCD contributions,''
\href{http://arXiv.org/abs/arXiv:0705.0746 [hep-ph]}{{\tt arXiv:0705.0746
  [hep-ph]}}.
%%CITATION = ARXIV:0705.0746;%%.

\bibitem{Okada:1990vk}
Y.~Okada, M.~Yamaguchi, and T.~Yanagida, ``Upper bound of the lightest Higgs
  boson mass in the minimal supersymmetric standard model,'' {\em Prog. Theor.
  Phys.} {\bf 85} (1991)
1--6.
%%CITATION = PTPKA,85,1;%%.

\bibitem{Ellis:1990nz}
J.~R. Ellis, G.~Ridolfi, and F.~Zwirner, ``Radiative corrections to the masses
  of supersymmetric Higgs bosons,'' {\em Phys. Lett.} {\bf B257} (1991)
83--91.
%%CITATION = PHLTA,B257,83;%%.

\bibitem{Haber:1990aw}
H.~E. Haber and R.~Hempfling, ``Can the mass of the lightest Higgs boson of the
  minimal supersymmetric model be larger than m(Z)?,'' {\em Phys. Rev. Lett.}
  {\bf 66} (1991)
1815--1818.
%%CITATION = PRLTA,66,1815;%%.

\bibitem{Haber:1996fp}
H.~E. Haber, R.~Hempfling, and A.~H. Hoang, ``Approximating the radiatively
  corrected {H}iggs mass in the minimal supersymmetric model,'' {\em Z.
  {P}hys.} {\bf C75} (1997) 539--554,
\href{http://arXiv.org/abs/hep-ph/9609331}{{\tt hep-ph/9609331}}.
%%CITATION = HEP-PH 9609331;%%.

\bibitem{Carena:1995bx}
M.~Carena, J.~R. Espinosa, M.~Quiros, and C.~E.~M. Wagner, ``Analytical
  expressions for radiatively corrected Higgs masses and couplings in the
  MSSM,'' {\em Phys. Lett.} {\bf B355} (1995) 209--221,
\href{http://arXiv.org/abs/hep-ph/9504316}{{\tt hep-ph/9504316}}.
%%CITATION = HEP-PH/9504316;%%.

\bibitem{Carena:1995wu}
M.~Carena, M.~Quiros, and C.~E.~M. Wagner, ``Effective potential methods and
  the Higgs mass spectrum in the MSSM,'' {\em Nucl. Phys.} {\bf B461} (1996)
  407--436,
\href{http://arXiv.org/abs/hep-ph/9508343}{{\tt hep-ph/9508343}}.
%%CITATION = HEP-PH/9508343;%%.

\bibitem{Essig:2007vq}
R.~Essig, ``Implications of the LEP Higgs bounds for the MSSM stop sector,''
\href{http://arXiv.org/abs/hep-ph/0702104}{{\tt hep-ph/0702104}}.
%%CITATION = HEP-PH 0702104;%%.

\bibitem{Ibanez:1983di}
L.~E. Ibanez and C.~Lopez, ``N=1 Supergravity, the Weak Scale and the
  Low-Energy Particle Spectrum,'' {\em Nucl. Phys.} {\bf B233} (1984)
511.
%%CITATION = NUPHA,B233,511;%%.

\bibitem{Carena:1996km}
M.~Carena, P.~H. Chankowski, M.~Olechowski, S.~Pokorski, and C.~E.~M. Wagner,
  ``Bottom-up approach and supersymmetry breaking,'' {\em Nucl. Phys.} {\bf
  B491} (1997) 103--128,
\href{http://arXiv.org/abs/hep-ph/9612261}{{\tt hep-ph/9612261}}.
%%CITATION = HEP-PH/9612261;%%.

\bibitem{Athron:2007as}
P.~Athron and D.~J. Miller, ``Fine Tuning in Supersymmetric Models,''
\href{http://arXiv.org/abs/arXiv:0707.1255 [hep-ph]}{{\tt arXiv:0707.1255
  [hep-ph]}}.
%%CITATION = ARXIV:0707.1255;%%.

\bibitem{Lahanas:1986uc}
A.~B. Lahanas and D.~V. Nanopoulos, ``The Road to No Scale Supergravity,'' {\em
  Phys. Rept.} {\bf 145} (1987)
1.
%%CITATION = PRPLC,145,1;%%.

\bibitem{TevatronTop:2007bx}
{\bf CDF} Collaboration, ``A combination of CDF and D0 results on the mass of
  the top quark,''
\href{http://arXiv.org/abs/hep-ex/0703034}{{\tt hep-ex/0703034}}.
%%CITATION = HEP-EX/0703034;%%.

\bibitem{PDBook}
W.-M. Yao {\em et al.}, ``{Review of Particle Physics},'' {\em {Journal of
  Physics G}} {\bf 33} (2006) 1+.

\bibitem{Sparticles:2004}
M.~Drees, R.~Godbole, and P.~Roy, {\em Theory \& Phenomenology of Sparticles}.
\newblock World Scientific Publishing Company, 2004.

\bibitem{Carena:2002es}
M.~Carena and H.~E. Haber, ``Higgs boson theory and phenomenology. (({V})),''
  {\em Prog. {P}art. {N}ucl. {P}hys.} {\bf 50} (2003) 63--152,
\href{http://arXiv.org/abs/hep-ph/0208209}{{\tt hep-ph/0208209}}.
%%CITATION = HEP-PH 0208209;%%.

\bibitem{Martin:1997ns}
S.~P. Martin, ``A supersymmetry primer,''
\href{http://arXiv.org/abs/hep-ph/9709356}{{\tt hep-ph/9709356}}.
%%CITATION = HEP-PH 9709356;%%.

\bibitem{Heinemeyer:1999be}
S.~Heinemeyer, W.~Hollik, and G.~Weiglein, ``The mass of the lightest MSSM
  Higgs boson: A compact analytical expression at the two-loop level,'' {\em
  Phys. Lett.} {\bf B455} (1999) 179--191,
\href{http://arXiv.org/abs/hep-ph/9903404}{{\tt hep-ph/9903404}}.
%%CITATION = HEP-PH 9903404;%%.

\bibitem{Kane:2004tk}
G.~L. Kane, T.~T. Wang, B.~D. Nelson, and L.-T. Wang, ``Theoretical
  implications of the LEP Higgs search,'' {\em Phys. Rev.} {\bf D71} (2005)
  035006,
\href{http://arXiv.org/abs/hep-ph/0407001}{{\tt hep-ph/0407001}}.
%%CITATION = HEP-PH 0407001;%%.

\bibitem{Kitano:2006gv}
R.~Kitano and Y.~Nomura, ``Supersymmetry, naturalness, and signatures at the
  LHC,'' {\em Phys. Rev.} {\bf D73} (2006) 095004,
\href{http://arXiv.org/abs/hep-ph/0602096}{{\tt hep-ph/0602096}}.
%%CITATION = HEP-PH 0602096;%%.

\bibitem{Ferreira:1995sn}
P.~M. Ferreira, I.~Jack, and D.~R.~T. Jones, ``Infrared soft universality,''
  {\em Phys. Lett.} {\bf B357} (1995) 359--364,
\href{http://arXiv.org/abs/hep-ph/9506467}{{\tt hep-ph/9506467}}.
%%CITATION = HEP-PH 9506467;%%.

\bibitem{Lanzagorta:1995ai}
M.~Lanzagorta and G.~G. Ross, ``Infrared fixed point structure of soft
  supersymmetry breaking mass terms,'' {\em Phys. Lett.} {\bf B364} (1995)
  163--174,
\href{http://arXiv.org/abs/hep-ph/9507366}{{\tt hep-ph/9507366}}.
%%CITATION = HEP-PH 9507366;%%.

\bibitem{Pendleton:1980as}
B.~Pendleton and G.~G. Ross, ``Mass and Mixing Angle Predictions from Infrared
  Fixed Points,'' {\em Phys. Lett.} {\bf B98} (1981)
291.
%%CITATION = PHLTA,B98,291;%%.

\bibitem{Frere:1983ag}
J.~M. Frere, D.~R.~T. Jones, and S.~Raby, ``Fermion Masses and Induction of the
  Weak Scale by Supergravity,'' {\em Nucl. Phys.} {\bf B222} (1983)
11.
%%CITATION = NUPHA,B222,11;%%.

\bibitem{Derendinger:1983bz}
J.~P. Derendinger and C.~A. Savoy, ``Quantum Effects and SU(2) x U(1) Breaking
  in Supergravity Gauge Theories,'' {\em Nucl. Phys.} {\bf B237} (1984)
307.
%%CITATION = NUPHA,B237,307;%%.

\bibitem{Gunion:1987qv}
J.~F. Gunion, H.~E. Haber, and M.~Sher, ``Charge / {C}olor {B}reaking {M}inima
  {A}nd {A}-{P}arameter {B}ounds {I}n {S}upersymmetric {M}odels,'' {\em Nucl.
  {P}hys.} {\bf B306} (1988)
1.
%%CITATION = NUPHA,B306,1;%%.

\bibitem{Casas:1995pd}
J.~A. Casas, A.~Lleyda, and C.~Munoz, ``Strong constraints on the parameter
  space of the MSSM from charge and color breaking minima,'' {\em Nucl. Phys.}
  {\bf B471} (1996) 3--58,
\href{http://arXiv.org/abs/hep-ph/9507294}{{\tt hep-ph/9507294}}.
%%CITATION = HEP-PH/9507294;%%.

\bibitem{LeMouel:2001ym}
C.~Le~Mouel, ``Charge and color breaking conditions associated to the top quark
  Yukawa coupling,'' {\em Phys. Rev.} {\bf D64} (2001) 075009,
\href{http://arXiv.org/abs/hep-ph/0103341}{{\tt hep-ph/0103341}}.
%%CITATION = HEP-PH/0103341;%%.

\bibitem{Kusenko:1996jn}
A.~Kusenko, P.~Langacker, and G.~Segre, ``Phase Transitions and Vacuum
  Tunneling Into Charge and Color Breaking Minima in the MSSM,'' {\em Phys.
  Rev.} {\bf D54} (1996) 5824--5834,
\href{http://arXiv.org/abs/hep-ph/9602414}{{\tt hep-ph/9602414}}.
%%CITATION = HEP-PH/9602414;%%.

\bibitem{Riotto:1995am}
A.~Riotto and E.~Roulet, ``Vacuum decay along supersymmetric flat directions,''
  {\em Phys. Lett.} {\bf B377} (1996) 60--66,
\href{http://arXiv.org/abs/hep-ph/9512401}{{\tt hep-ph/9512401}}.
%%CITATION = HEP-PH/9512401;%%.

\bibitem{Dermisek:2006qj}
R.~Dermisek, H.~D. Kim, and I.-W. Kim, ``Mediation of supersymmetry breaking in
  gauge messenger models,'' {\em JHEP} {\bf 10} (2006) 001,
\href{http://arXiv.org/abs/hep-ph/0607169}{{\tt hep-ph/0607169}}.
%%CITATION = HEP-PH/0607169;%%.

\bibitem{Culbertson:2000am}
{\bf SUSY Working Group} Collaboration, R.~Culbertson {\em et al.}, ``Low-scale
  and gauge-mediated supersymmetry breaking at the Fermilab Tevatron Run II,''
\href{http://arXiv.org/abs/hep-ph/0008070}{{\tt hep-ph/0008070}}.
%%CITATION = HEP-PH/0008070;%%.

\bibitem{Chou:1999zb}
C.-L. Chou and M.~E. Peskin, ``Scalar top quark as the next-to-lightest
  supersymmetric particle,'' {\em Phys. Rev.} {\bf D61} (2000) 055004,
\href{http://arXiv.org/abs/hep-ph/9909536}{{\tt hep-ph/9909536}}.
%%CITATION = HEP-PH/9909536;%%.

\bibitem{Sarid:1999zx}
U.~Sarid and S.~D. Thomas, ``Mesino-antimesino oscillations,'' {\em Phys. Rev.
  Lett.} {\bf 85} (2000) 1178--1181,
\href{http://arXiv.org/abs/hep-ph/9909349}{{\tt hep-ph/9909349}}.
%%CITATION = HEP-PH/9909349;%%.

\bibitem{Perelstein:2007nx}
M.~Perelstein and C.~Spethmann, ``A collider signature of the supersymmetric
  golden region,'' {\em JHEP} {\bf 04} (2007) 070,
\href{http://arXiv.org/abs/hep-ph/0702038}{{\tt hep-ph/0702038}}.
%%CITATION = HEP-PH/0702038;%%.

\bibitem{Dermisek:2007fi}
R.~Dermisek and I.~Low, ``Probing the stop sector and the sanity of the MSSM
  with the Higgs boson at the LHC,''
\href{http://arXiv.org/abs/hep-ph/0701235}{{\tt hep-ph/0701235}}.
%%CITATION = HEP-PH 0701235;%%.

\bibitem{Espinosa:1999zm}
J.~R. Espinosa and R.-J. Zhang, ``MSSM lightest CP-even Higgs boson mass to
  O(alpha(s) alpha(t)): The effective potential approach,'' {\em JHEP} {\bf 03}
  (2000) 026,
\href{http://arXiv.org/abs/hep-ph/9912236}{{\tt hep-ph/9912236}}.
%%CITATION = HEP-PH/9912236;%%.

\bibitem{Carena:2000dp}
M.~Carena {\em et al.}, ``Reconciling the two-loop diagrammatic and effective
  field theory computations of the mass of the lightest CP-even Higgs boson in
  the MSSM,'' {\em Nucl. Phys.} {\bf B580} (2000) 29--57,
\href{http://arXiv.org/abs/hep-ph/0001002}{{\tt hep-ph/0001002}}.
%%CITATION = HEP-PH 0001002;%%.

\bibitem{Heinemeyer:2004ms}
S.~Heinemeyer, ``M{SSM} {H}iggs physics at higher orders,''
\href{http://arXiv.org/abs/hep-ph/0407244}{{\tt hep-ph/0407244}}.
%%CITATION = HEP-PH 0407244;%%.

\bibitem{Carena:1999xa}
M.~Carena, S.~Heinemeyer, C.~E.~M. Wagner, and G.~Weiglein, ``Suggestions for
  improved benchmark scenarios for Higgs- boson searches at LEP2,''
\href{http://arXiv.org/abs/hep-ph/9912223}{{\tt hep-ph/9912223}}.
%%CITATION = HEP-PH 9912223;%%.

\bibitem{Martin:1993zk}
S.~P. Martin and M.~T. Vaughn, ``Two loop renormalization group equations for
  soft supersymmetry breaking couplings,'' {\em Phys. Rev.} {\bf D50} (1994)
  2282,
\href{http://arXiv.org/abs/hep-ph/9311340}{{\tt hep-ph/9311340}}.
%%CITATION = HEP-PH/9311340;%%.

\bibitem{Hill:1980sq}
C.~T. Hill, ``Quark and Lepton Masses from Renormalization Group Fixed
  Points,'' {\em Phys. Rev.} {\bf D24} (1981)
691.
%%CITATION = PHRVA,D24,691;%%.

\bibitem{Lanzagorta:1995gp}
M.~Lanzagorta and G.~G. Ross, ``Infrared fixed points revisited,'' {\em Phys.
  Lett.} {\bf B349} (1995) 319--328,
\href{http://arXiv.org/abs/hep-ph/9501394}{{\tt hep-ph/9501394}}.
%%CITATION = HEP-PH/9501394;%%.

\end{thebibliography}\endgroup
}}

\end{document}